\documentclass{aa} %
\usepackage[varg]{txfonts}
\usepackage{amsmath}
\usepackage{amssymb}
\usepackage{graphicx}
\usepackage{booktabs}
\usepackage{marvosym}
\usepackage{txfonts}
\usepackage{multirow}
\usepackage{rotating}
\usepackage{longtable}
\usepackage{lscape}
\usepackage{epstopdf}
\usepackage{hyperref}
\usepackage[svgnames]{xcolor}
\hypersetup{colorlinks,citecolor=Blue}

\def\f#1   {Fig.~\ref{#1}}
\def\s#1   {Sect.~\ref{#1}}
\def\tab#1   {Tab.~\ref{#1}}
\def\t#1   {Tab.~\ref{#1}}
\def\eq#1   {Eq.~\ref{#1}}
\def\lum   {$\mathrm{L}_\mathrm{1.4GHz}$}
\def\comm#1   {{\tt (COMMENT: #1) }}

\def\sqdeg            { square degree} 

\def\msol              {$\mathrm{M}_{\odot}$}

\def\msun              {$\mathrm{M}_{\odot}$}

\def\wh                {W~Hz$^{-1}$}

\def\mic               {$\mathrm{\mu m}$}

\def\nsource   {10,830}

\begin{document}

\authorrunning{Smol\v{c}i\'{c} et al.}
\titlerunning{Observational constraints on radio-mode AGN feedback out to $z\sim5$}

\title{The VLA-COSMOS 3~GHz Large Project: \\ Cosmic evolution of radio AGN and implications for radio-mode feedback since $z\sim5$}

\author{
        V.~Smol\v{c}i\'{c}\inst{1}, 
        M.~Novak\inst{1}, 
I.~Delvecchio\inst{1},
        L.~Ceraj\inst{1}, 
        M.~Bondi\inst{2},
        J.~Delhaize\inst{1},
S.~Marchesi\inst{3},
E.~Murphy\inst{4},
        E.~Schinnerer\inst{5},
        E.~Vardoulaki\inst{6},
 G.~Zamorani\inst{7} 
                }
\institute{
Department of Physics, Faculty of Science, University of Zagreb,  Bijeni\v{c}ka cesta 32, 10000  Zagreb, Croatia
\and
Istituto di Radioastronomia di Bologna - INAF, via P. Gobetti, 101, 40129, Bologna, Italy 
\and
Department of Physics \& Astronomy, Clemson University, Clemson, SC 29634, USA
\and
National Radio Astronomy Observatory, 520 Edgemont Road, Charlottesville, VA 22903, USA
\and
Max-Planck-Institut f$\ddot{u}$r Astronomie, K$\ddot{o}$nigstuhl 17, D-69117 Heidelberg, Germany
\and
Argelander Institut for Astronomy, Auf dem H\"{u}gel 71, Bonn, 53121, Germany
\and
INAF-Osservatorio  Astronomico di Bologna, Via Piero Gobetti 93/3, I - 40129 Bologna, Italy
}

   \date{Received ; accepted}

\abstract{
Based on a sample of over $1,800$ radio AGN at redshifts out to $z\sim5,$ which have typical stellar masses within $\sim3\times(10^{10}-10^{11})$~\msol,  and 
3\,GHz radio data in the COSMOS field, we derived the 1.4\,GHz radio luminosity functions for  radio AGN (\lum~$\sim10^{22}-10^{27}$~\wh )  out to $z\sim5$. We constrained the evolution of this population via continuous models of pure density and pure luminosity 
evolutions, and we found best-fit parametrizations of $\Phi^*\propto(1+z)^{(2.00\pm0.18)-(0.60\pm0.14)z}$, and $L^*\propto(1+z)^{(2.88\pm0.82)-(0.84\pm0.34)z}$, respectively, with a turnover in number and luminosity densities of the population at $z\approx1.5$. We converted 1.4\,GHz luminosity to kinetic luminosity taking uncertainties of the scaling relation used into account. We thereby derived the cosmic evolution of the kinetic luminosity density provided by the AGN and compared this luminosity density to the radio-mode AGN feedback assumed in the Semi-Analytic Galaxy Evolution (SAGE) model, i.e., to the redshift evolution of the central supermassive black hole accretion luminosity taken in the model as the source of heating that offsets the energy losses of the cooling, hot halo gas, and thereby limits further stellar mass growth of massive galaxies. 
We find that the kinetic luminosity exerted by our radio AGN may be high enough to balance the radiative cooling of the hot gas at 
each cosmic epoch since $z\sim5$. However, although our findings  support the idea of radio-mode AGN feedback as a cosmologically relevant process in massive galaxy formation, many simplifications in both the observational and semi-analytic approaches still remain and need to be resolved before robust conclusions can be reached. 
}


\maketitle

\section{Introduction}
\label{sec:intro}

Understanding massive galaxy formation is one of the major quests of modern astrophysics, and optimally studied via synergy between theory (magnetohydrodynamic and semi-analytic simulations) and (panchromatic) observations. Our current understanding of massive galaxy formation requires energetic outflows from radio-luminous active galactic nuclei (AGN) to suppress stellar growth in the most massive galaxies \citep[e.g.,][]{benson03,bower06,croton06,croton16}. This is mostly based on the advance of semi-analytic models in the last decades.

Semi-analytic models are used to  study the properties of galaxies formed in cosmological models within dark matter halos by
describing ongoing physical processes, such as gas cooling, star formation, merger rates, and various types of feedback. One of their major past challenges was related to substantial overprediction of the quantity of the most massive galaxies in the universe \citep[see, e.g.,][]{steinmetz97}. This was  solved by introducing a process, dubbed radio-mode AGN feedback \citep[e.g.,][]{croton06}.

By now radio-mode AGN feedback has become a standard and key ingredient in semi-analytic models that enables reproduction of the observed galaxy properties well \citep{croton06, croton16, bower06, sijacki07}. The feedback is assumed to be related to radio AGN outflows  as the main source that heats the  halo gas surrounding a massive galaxy, thereby quenching its star formation and limiting
growth, and thus avoids creating overly  massive galaxies.  
Introducing  such a feedback process was motivated by radio AGN that were observed to i)
have massive galaxy hosts \citep[e.g.,][]{best05, kauffmann08, smo09a} and
ii) interact with, i.e.,\ heat the intra-cluster medium on large (group/cluster) scales, potentially solving
the so-called cooling-flow problem in galaxy clusters (see \citealt{fabian12} for a review). 

Initial implementations of radio-mode feedback were aimed to solve the cooling-flow problem and generalized based on simple phenomenological descriptions \citep{croton06} or a sharp cutoff in cooling \citep{bower06} that were both applied beyond a critical halo mass threshold. For example, in the \citet{croton06} model the source of heating was related to low-luminosity radio activity caused by hot gas accretion onto the central supermassive black hole (SMBH), once a static hot halo had formed around the host galaxy. The halo virial mass threshold beyond which a static hot halo forms was taken to be $\gtrsim2.5\cdot10^{11}$~\msol \ (see their Fig.~2). 

A more complex cycle between gas cooling and radio-mode AGN heating was recently implemented within the updated \citet{croton06} model, i.e., the Semi-Analytic Galaxy Evolution (SAGE) model \citep{croton16}.  In this model cooling and heating of the halo gas has been more directly coupled and the phenomenological treatment of radio-mode feedback has been put on more physical grounds by assuming  accretion onto central SMBHs hosted by  massive haloes of spherical, Bondi-Hoyle type \citep{bondi52},  and has been scaled by a radio-mode efficiency parameter that modulates the strength of the accretion and subsequently the radio-mode feedback. 

From an observational point of view, radio-mode AGN feedback, and particularly its  cosmological relevance for massive galaxy formation, as suggested by the semi-analytic models, is difficult and challenging to test. While observations of large, spatially resolved radio galaxies inducing cavities (i.e., buoyantly rising bubbles) in the hot X-ray emitting intra-cluster medium (ICM) are possible for a small number of well-studied, nearby systems \citep[e.g.,][]{birzan04, birzan08, osullivan11}, at higher redshifts the observational data is limited as most of the systems are unresolved in both the radio and X-ray bands. This necessitates the use of scaling relations to convert the monochromatic radio luminosity to kinetic luminosity inferred either on the basis of well-studied nearby systems \citep{merloni07, birzan08,cavagnolo10,osullivan11, godfrey16} and assumed to  hold at high redshifts \citep{smo09a,lafranca10,pracy16}, or on theoretical grounds, including various assumptions drawn from well-studied and resolved radio galaxies \citep{willott01, daly12, godfrey16}, and again assumed to hold for populations of radio AGN that have been observed in deep radio surveys over a large redshift range \citep{best06,best14}. 

A commonly used approach  to estimate  the cosmic evolution of radio-mode feedback, i.e.,\ that of the volume averaged kinetic luminosity density of radio AGN, is based on constraining the cosmic evolution of radio AGN luminosity functions, convolved with a scaling relation between the monochromatic  and kinetic luminosities \citep{merloni08, cattaneo09, smo09a, smo15, best14, pracy16}. Hence, the cosmic evolution of radio AGN has direct implications on constraining the redshift evolution of radio-mode AGN feedback.

Past studies have shown that radio AGN evolve differentially, consistent with cosmic downsizing; low radio-luminosity sources
evolve less strongly than high radio-luminosity sources and the number density peak occurs at higher redshift for higher luminosity radio AGN 
\citep{dunlop90, willott01, waddington01, rigby11}.
Consistent with the evolution of optically and X-ray selected quasars
\citep[e.g.,][]{schmidt95,silverman08,brusa09}, high-luminosity radio AGN (\lum~$> 2\times10^{26}$~\wh ) show
a strong positive density evolution with redshift out to $z\sim 2$,
beyond which their comoving volume density starts decreasing \citep{dunlop90, willott01}.  Intermediate- and low-luminosity radio AGN
(\lum~$\sim1.6\times10^{24} - 3\times10^{26}$~\wh ) have been shown to evolve slower
with a comoving volume density turnover occurring at a lower redshift
($z \sim 1 - 1.5$; \citealt{waddington01, clewley04, sadler07, donoso09, smo09a}). 

The  goal of recent studies has been to constrain the cosmic evolution of the low-luminosity radio AGN reaching toward the epoch of reionization \citep[$z\lesssim6$; e.g.,][]{mcalpine13,smo15,padovani15}. This is  becoming possible only now via deep panchromatic surveys, including radio data obtained with the recently upgraded radio facilities, such as the Karl. G. Jansky Very Large Array (VLA). 
In this context, the VLA-COSMOS 3 GHz Large Project \citep{smo17a} is to date the deepest radio continuum survey over a relatively large, 2 square degree field. Combined with the rich COSMOS multiwavelength data \citep{scoville07}, the Large Project thus provides a unique data set to study the evolution of the faintest observable radio AGN since $z\sim5$ to date  and puts the results in  context of  radio-mode AGN feedback as an assumed cosmologically relevant process for massive galaxy formation. This is the aim of the work presented here.

In \s{sec:data} \ we describe the VLA-COSMOS 1.4 and 3 GHz projects and the 3~GHz selected radio AGN sample used here. In \s{sec:lf} \ we derive the 1.4~GHz rest-frame radio luminosity functions for our AGN and model their evolution out to $z\sim5$. In \s{sec:feedback} \ we consider the obtained results in the context of radio-mode AGN feedback, comparing them with the SAGE semi-analytic model, and in \s{sec:unknowns} \ we discuss various unknowns remaining in both, the observational and analytic approaches. We summarize in \s{sec:summary} . Throughout the paper we adopt $H_0=70,\, \Omega_M=0.3, \Omega_\Lambda = 0.7$. We define the radio spectral index, $\alpha$, via $S_\nu\propto\nu^\alpha$, where $S_\nu$ is the flux density at frequency $\nu$. We use a \citet{chabrier03} initial mass function.

\section{Data and radio AGN sample}
\label{sec:data}

For our analysis we employ the VLA-COSMOS 3~GHz Large Project \citep{smo17a} and the VLA-COSMOS 1.4~GHz Large and Deep Projects \citep{schinnerer04,schinnerer07,schinnerer10}. The VLA-COSMOS 3~GHz Large Project provides radio data at 10~cm wavelength within the COSMOS 2\sqdeg \ field down to an average ($1\sigma$) sensitivity of $2.3~\mu$Jy/beam over a $0.75\arcsec$ resolution element. The source catalog lists \nsource \ sources with signal-to-noise ratios (S/N) $\geq5$ (see \citealt{smo17a} for details). 
Within the VLA-COSMOS 1.4~GHz  Large Project the 2\sqdeg\ COSMOS field has been observed with the VLA
down to a uniform $rms$ of $\sim10-15~\mu$Jy/beam at a resolution of
$1.5\arcsec$. The VLA-COSMOS Deep Project has added further VLA 1.4~GHz observations
to the inner square degree yielding an $rms$  in this area of $\sim7\,(12)~\mu$Jy/beam over a $1.5\arcsec\,(2.5\arcsec)$ resolution element. The joint catalog lists 2,864 sources with S/N ratios  $\geq5$ at  $1.5\arcsec$ and/or $2.5\arcsec$ resolution in the Large and/or Deep Projects. The number of sources detected in both the 3 and 1.4~GHz catalogs is 2,530  (see \citealt{smo17a}).

Multiwavelength counterparts of the VLA-COSMOS 3~GHz Large Project sources have been identified by \citet{smo17b}. 
For the cross-correlation they used the i) most up-to-date COSMOS photometric redshift catalog (COSMOS2015 hereafter; see \citealt{laigle16} for details),
ii)  i-band selected catalog \citep{capak07}, and iii)  3.6~\mic \ selected catalog \citep{sanders07}. The latter two were used to supplement potentially missed counterparts due to a nondetection in the COSMOS2015 catalog. To assure the most reliable photometric redshifts and a clean selection function here we used the most secure counterparts identified in the COSMOS2015 catalog within an area of 1.77 square degrees, uncontaminated by bright stars or saturated objects (see also Delvecchio et al.\ 2017; Delhaize et al.\ 2017; and \citealt{novak17}). Within this area 93\% (8,035/8696) of the 3~GHz radio sources were associated with (COSMOS2015, $i$-band or IRAC) counterparts;  96\% (7,729) of these have counterparts drawn from the COSMOS2015 catalog and, thus, are the most precise photometric redshifts available (see \citealt{laigle16}; see also Fig.~4 in \citealt{smo17b}).

Using the full COSMOS (ultraviolet to millimeter) multiwavelength data set a three-component SED fitting procedure was performed for each galaxy taking into account the energy balance between the ultraviolet (UV) and infrared (IR) emissions of the galaxy and the contribution from a potential AGN component (see Delvecchio et al.\ 2017 for details; see also \citealt{smo17b}). These fits yielded an estimate of the total IR luminosity that was associated with the best-fit star-forming galaxy template (i.e., the AGN contribution was subtracted, if present),  which was then converted to a star formation rate via the \citet{kennicutt98} conversion, assuming a \citet{chabrier03} initial mass function. To identify galaxies whose radio emission predominantly arises from the AGN,  galaxies with radio excess were identified via a redshift dependent ($3\sigma$) threshold in the distribution of the ratio of the logarithms of radio luminosity (which is directly proportional to star-formation rate; e.g., \citealt{condon92}) over the IR-derived star formation rate  (see Fig.~5 in \citealt{smo17b}). 

We define our radio AGN sample as the  sample of 3~GHz sources with   COSMOS2015 counterparts, with photometric or spectroscopic redshifts, detected in the  near-infrared
(NIR), which exhibit radio emission in ($>3\sigma$) excess of that expected from their hosts' (IR-based) star formation rates. The sample contains 1,814 sources and the chosen criterion assures that at least 80\% of their radio emission is due to the AGN component. About 50\% (899) of these are also detected at 1.4~GHz, allowing us to derive their spectral indices directly. In \f{fig:mstar} \ we show the redshift distribution of our radio AGN sample, the 1.4~GHz rest-frame radio luminosity and stellar mass as a function of redshift. The sample reaches out to a redshift of $z\sim5$ and comprises 1.4 GHz radio luminosities in the range of $10^{22}-10^{27}$~\wh  and typical stellar masses within $\sim3\times(10^{10}-10^{11})$~\msol . 
About 3\% of the sources have estimated stellar masses that are lower than $\sim3\times10^{9}$~\msun . Although radio selection and X-ray selection preferentially select high-mass galaxies (e.g., \citealt{smo09,smo09a}), this fraction is very similar to the fraction of low-mass galaxies in samples of X-ray selected AGN (see, for example, Fig. 1 in \citealt{bongiorno16}, where the host galaxy mass function of the XMM-selected AGN in COSMOS is analyzed). The classification of these low-mass AGN as radio-excess sources seems to be statistically robust.

\begin{figure}
\includegraphics[bb= 0 0 550 339, width=\columnwidth]{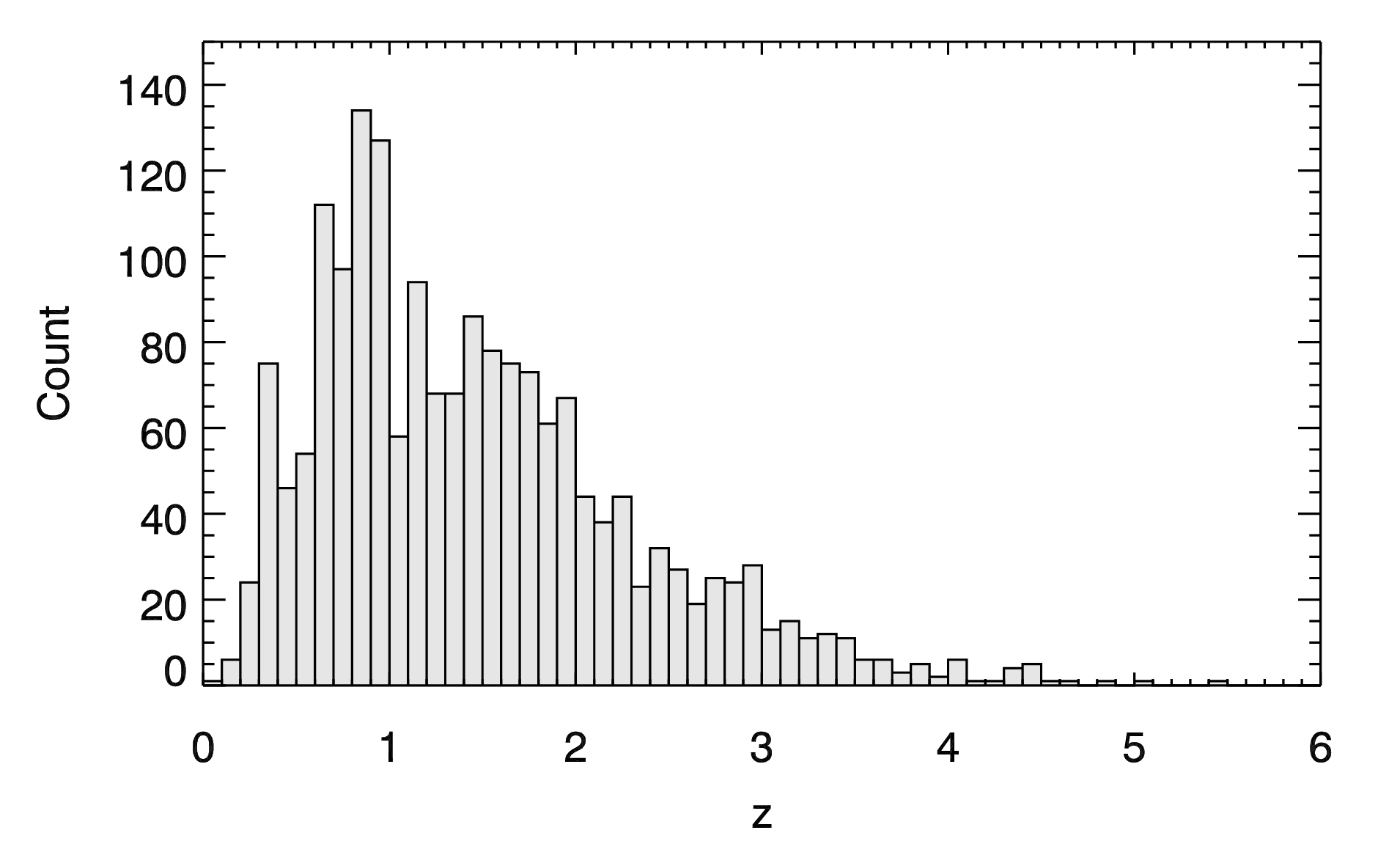}
\includegraphics[bb= 0 0 550 339, width=\columnwidth]{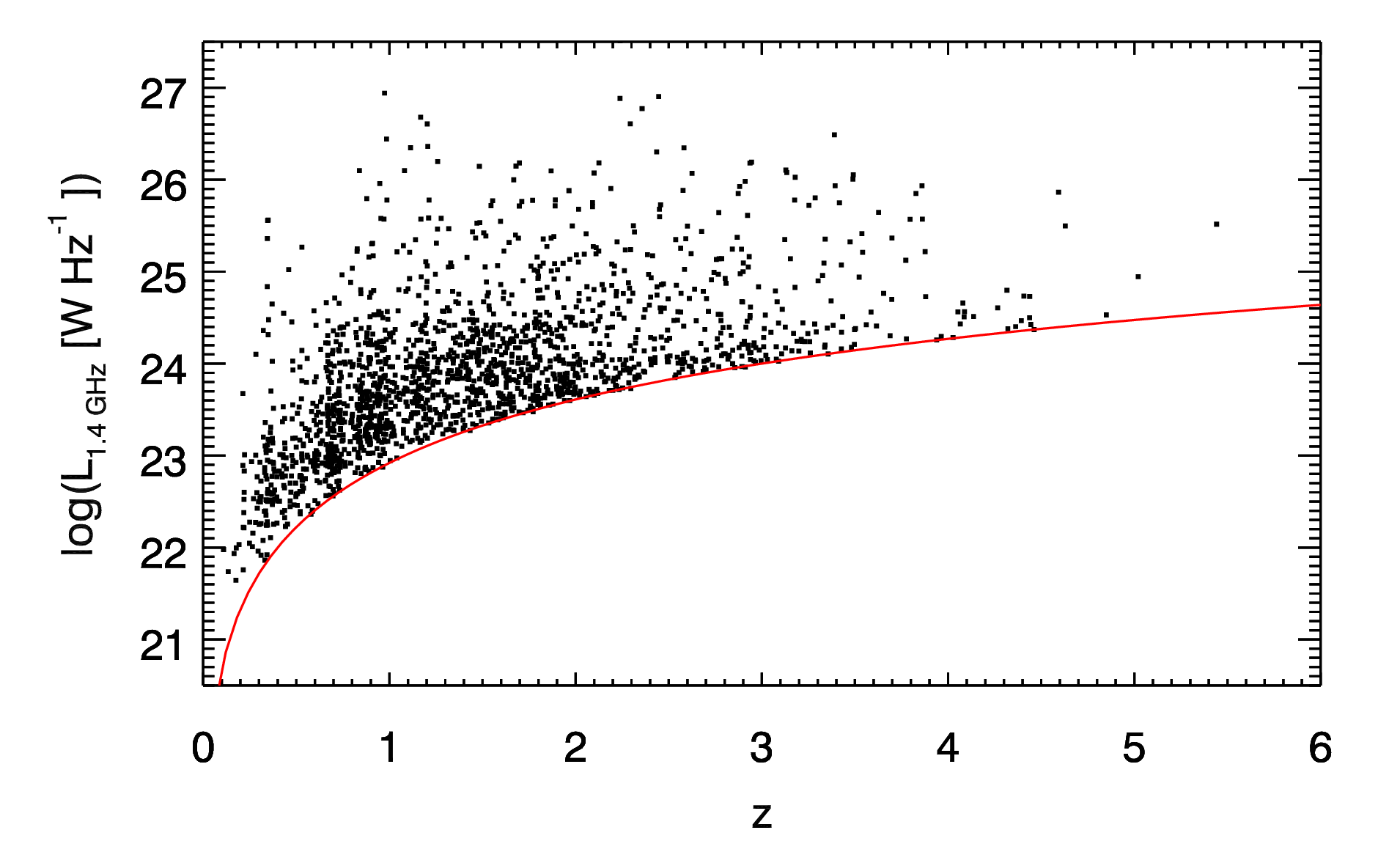}
\includegraphics[bb= 5 0 432 272, width=\columnwidth]{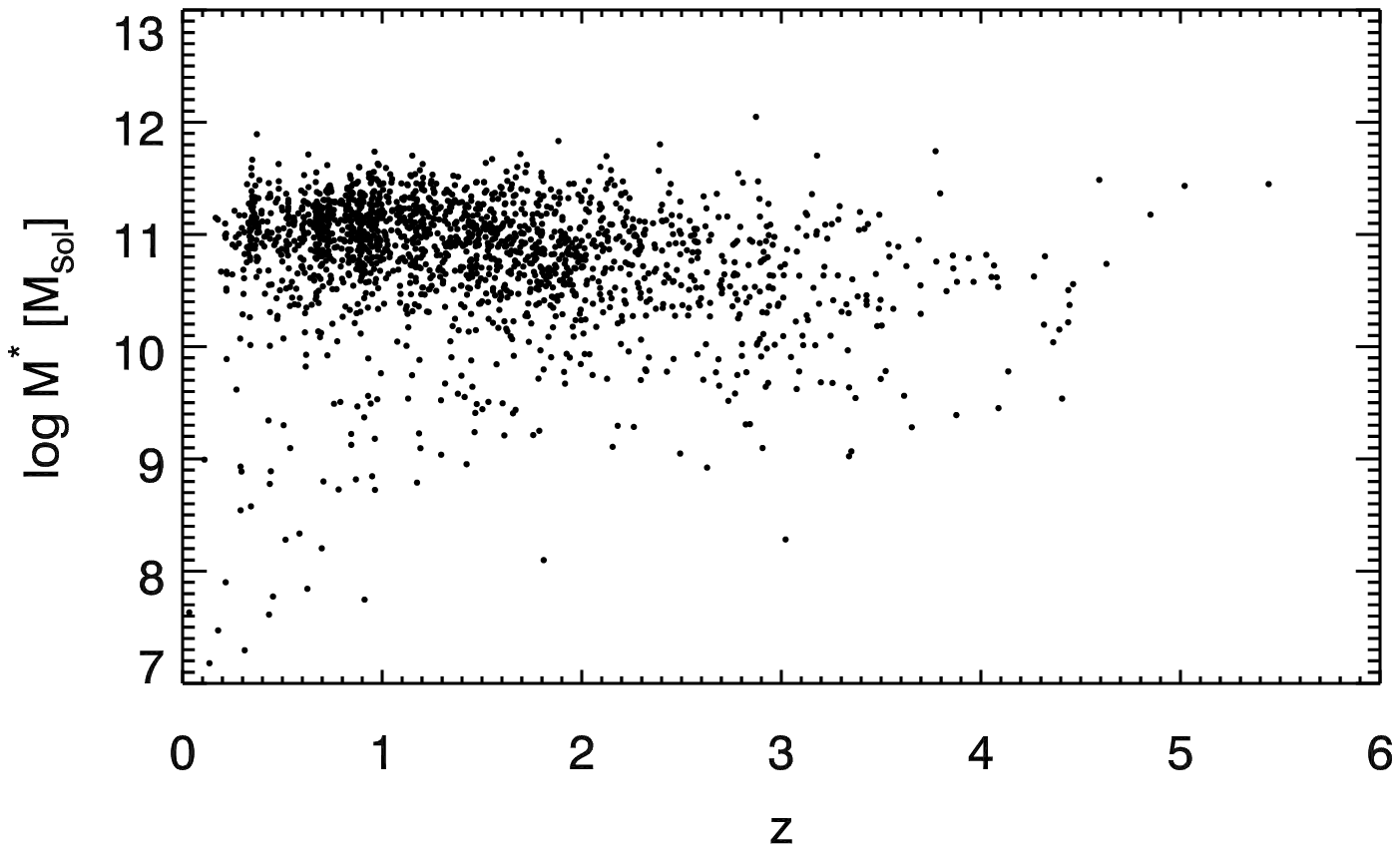}
\caption{Redshift distribution (top panel), 1.4~GHz rest-frame radio luminosity (middle panel), and stellar mass (bottom panel) as a function of redshift for our radio-excess AGN sample. 
  \label{fig:mstar}}
\end{figure}

\section{Radio AGN luminosity functions and their cosmic evolution }
\label{sec:lf}

In this Section we derive the AGN luminosity functions out to $z\sim5$ (\s{sec:lfs} ) and model the cosmic evolution of radio AGN in the COSMOS field (\s{sec:lfevolv} ). We further present the cosmic evolution of their number and luminosity densities (\s{sec:ld} ) and a comparison with results from the literature (\s{sec:compare} ).

\subsection{Radio AGN luminosity functions out to $z\sim5$}
\label{sec:lfs}

We computed the volume densities (for a given redshift range) as described in detail by \citet{novak17}. We followed the $\mathrm{1/V_{max}}$ procedure and corrected for a combined set of radio incompletenesses, including radio detection, noise, and resolution biases, as well as for the incompleteness of the counterpart catalog due to radio sources without assigned NIR counterparts (the latter being overall less than 10\%). We refer to Sec. 3.1.\ in \citet{novak17} for a detailed description of the procedure (see also their Fig.~2). 

\begin{figure}
\includegraphics[width=\columnwidth]{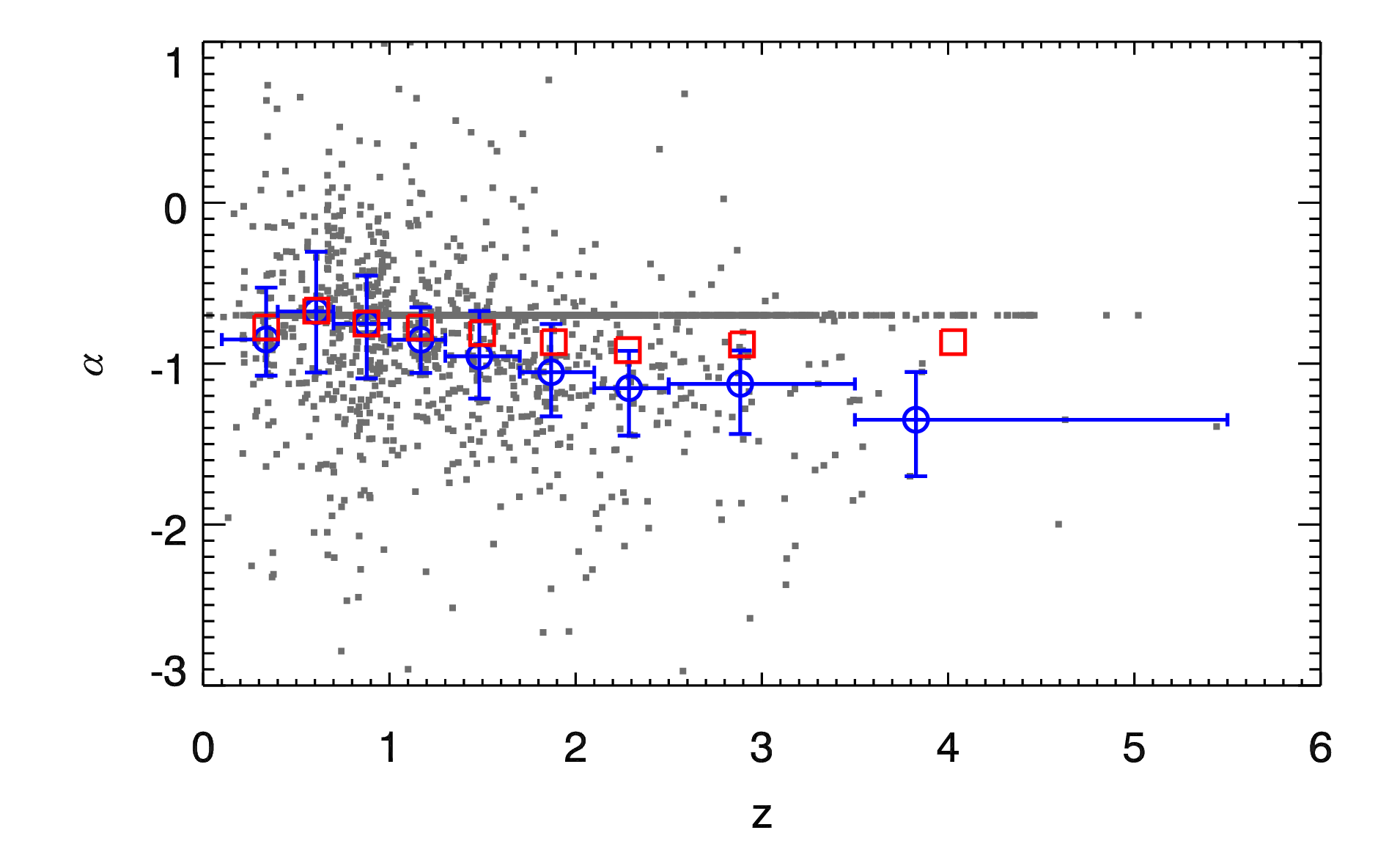}
\caption{Spectral indices of our radio-excess AGN sample as a function of redshift  (gray points). Median values with interquartile ranges for sources with measured 1.4-3~GHz spectral indices in different redshift bins are shown with blue circles. Mean values for all sources (i.e., also undetected at 1.4~GHz for which we manually set $\alpha=-0.7$) in same redshift bins are shown with red squares. 
  \label{fig:alphas}}
\end{figure}

The rest-frame 1.4~GHz luminosities\footnote{For simplicity and easier comparison with the literature, we here derive radio luminosities at 1.4 GHz rest-frame frequency. This corresponds to  the most commonly used reference frequency in the literature.}
 were computed using the observed-frame 3~GHz flux densities. If the source was also detected at 1.4~GHz  the inferred spectral index was used in the computation. If the source was undetected at 1.4~GHz we assumed a spectral index of $\alpha=-0.7$, which corresponds to the median value derived for all radio sources detected at 3~GHz and also taking the limits (i.e., nondetections at 1.4 GHz) into account via survival analysis (see also \citealt{smo17a}). Spectral indices used in the further analysis are shown in Fig.~\ref{fig:alphas}. A trend toward steeper spectra can be seen with increasing redshift. Without deep observations at other radio frequencies we cannot state whether this trend is real  (e.g., owing to spectral steepening of the radio spectrum at higher rest-frame frequencies, as sampled for our high-redshift sources) or a bias due to our flux limited observations.

The 1.4~GHz radio luminosity functions (LFs) for our radio AGN, separated into nine redshift ranges out to $z\sim5$,  are shown in \f{fig:lfs} , and tabulated in \t{tab:lumfun_vmax} . Redshift bins were chosen to be large enough to contain a statistically significant number of galaxies and to mitigate possible photometric uncertainties (i.e., sources falling into the wrong bin). We report LFs using the median luminosity in each luminosity bin, while the error bars show the width of the luminosity bin.

\begin{table*}
\begin{center}
\caption{Luminosity functions of radio-excess AGN obtained with the $V_\text{max}$ method.  The listed luminosity values represent the median luminosity of the sources in the corresponding luminosity bin. }
\renewcommand{\arraystretch}{1.5}
\begin{tabular}[t]{c c c c}
\hline
Redshift
 & $\log\left(\dfrac{L_{1.4\,\text{GHz}}}{\text{W}\,\text{Hz}^{-1}}\right)$
 & $\log\left(\dfrac{\Phi}{\text{Mpc}^{-3}\,\text{dex}^{-1}}\right)$
 & N \\
\hline 
$0.1<z<0.4$ & 21.92$_{-0.30}^{+0.081}$ & -3.83$_{-0.15}^{+0.16}$ & 10 \\
 & 22.48$_{-0.48}^{+0.24}$ & -3.94$_{-0.058}^{+0.067}$ & 49 \\
 & 22.93$_{-0.22}^{+0.50}$ & -4.15$_{-0.070}^{+0.083}$ & 33 \\
 & 23.70$_{-0.27}^{+0.45}$ & -4.97$_{-0.22}^{+0.25}$ & 5 \\
 & 24.48$_{-0.33}^{+0.39}$ & -4.97$_{-0.22}^{+0.25}$ & 5 \\
 & 25.56$_{-0.69}^{+0.031}$ & -5.19$_{-0.30}^{+0.34}$ & 3 \\
$0.4<z<0.7$ & 22.41$_{-0.20}^{+0.15}$ & -4.14$_{-0.097}^{+0.13}$ & 19 \\
 & 22.89$_{-0.33}^{+0.21}$ & -3.99$_{-0.042}^{+0.047}$ & 97 \\
 & 23.36$_{-0.26}^{+0.29}$ & -4.26$_{-0.053}^{+0.061}$ & 59 \\
 & 23.96$_{-0.30}^{+0.24}$ & -4.65$_{-0.079}^{+0.097}$ & 25 \\
 & 24.44$_{-0.24}^{+0.31}$ & -5.05$_{-0.15}^{+0.16}$ & 10 \\
 & 25.16$_{-0.42}^{+0.13}$ & -5.75$_{-0.37}^{+0.45}$ & 2 \\
$0.7<z<1.0$ & 22.86$_{-0.29}^{+0.064}$ & -4.11$_{-0.12}^{+0.17}$ & 25 \\
 & 23.34$_{-0.42}^{+0.39}$ & -4.07$_{-0.032}^{+0.035}$ & 208 \\
 & 24.05$_{-0.32}^{+0.49}$ & -4.42$_{-0.052}^{+0.059}$ & 98 \\
 & 24.80$_{-0.26}^{+0.55}$ & -5.19$_{-0.090}^{+0.11}$ & 19 \\
 & 25.79$_{-0.44}^{+0.37}$ & -5.69$_{-0.20}^{+0.22}$ & 6 \\
 & 26.76$_{-0.60}^{+0.21}$ & -6.17$_{-0.37}^{+0.45}$ & 2 \\
$1.0<z<1.3$ & 23.14$_{-0.21}^{+0.048}$ & -4.46$_{-0.17}^{+0.18}$ & 8 \\
 & 23.57$_{-0.39}^{+0.31}$ & -4.36$_{-0.039}^{+0.043}$ & 114 \\
 & 24.11$_{-0.22}^{+0.48}$ & -4.66$_{-0.049}^{+0.056}$ & 69 \\
 & 24.91$_{-0.32}^{+0.39}$ & -5.29$_{-0.097}^{+0.13}$ & 16 \\
 & 25.57$_{-0.27}^{+0.43}$ & -5.67$_{-0.19}^{+0.20}$ & 7 \\
 & 26.36$_{-0.35}^{+0.35}$ & -5.73$_{-0.20}^{+0.22}$ & 6 \\
$1.3<z<1.7$ & 23.39$_{-0.14}^{+0.063}$ & -4.22$_{-0.080}^{+0.097}$ & 30 \\
 & 23.78$_{-0.33}^{+0.23}$ & -4.40$_{-0.037}^{+0.040}$ & 132 \\
 & 24.22$_{-0.22}^{+0.33}$ & -4.56$_{-0.042}^{+0.046}$ & 101 \\
 & 24.74$_{-0.19}^{+0.36}$ & -5.21$_{-0.082}^{+0.10}$ & 24 \\
 & 25.40$_{-0.29}^{+0.26}$ & -5.49$_{-0.11}^{+0.14}$ & 13 \\
 & 26.00$_{-0.34}^{+0.21}$ & -5.77$_{-0.19}^{+0.20}$ & 7 \\
\hline
\end{tabular}
\quad
\renewcommand{\arraystretch}{1.5}
\begin{tabular}[t]{c c c c}
\hline
Redshift
 & $\log\left(\dfrac{L_{1.4\,\text{GHz}}}{\text{W}\,\text{Hz}^{-1}}\right)$
 & $\log\left(\dfrac{\Phi}{\text{Mpc}^{-3}\,\text{dex}^{-1}}\right)$
 & N \\
\hline  
$1.7<z<2.1$ & 23.60$_{-0.16}^{+0.060}$ & -4.14$_{-0.098}^{+0.13}$ & 30 \\
 & 23.86$_{-0.20}^{+0.29}$ & -4.41$_{-0.041}^{+0.045}$ & 106 \\
 & 24.39$_{-0.24}^{+0.25}$ & -4.75$_{-0.054}^{+0.061}$ & 60 \\
 & 24.92$_{-0.28}^{+0.22}$ & -5.13$_{-0.075}^{+0.091}$ & 28 \\
 & 25.32$_{-0.18}^{+0.31}$ & -5.56$_{-0.15}^{+0.16}$ & 10 \\
 & 25.77$_{-0.14}^{+0.35}$ & -5.53$_{-0.15}^{+0.16}$ & 10 \\
$2.1<z<2.5$ & 23.74$_{-0.083}^{+0.091}$ & -4.40$_{-0.13}^{+0.18}$ & 13 \\
 & 24.04$_{-0.21}^{+0.53}$ & -4.77$_{-0.049}^{+0.055}$ & 74 \\
 & 24.84$_{-0.27}^{+0.48}$ & -5.12$_{-0.069}^{+0.082}$ & 37 \\
 & 25.60$_{-0.29}^{+0.46}$ & -5.94$_{-0.19}^{+0.20}$ & 7 \\
 & 26.48$_{-0.43}^{+0.32}$ & -6.18$_{-0.25}^{+0.28}$ & 4 \\
 & 26.90$_{-0.11}^{+0.64}$ & -6.31$_{-0.30}^{+0.34}$ & 3 \\
$2.5<z<3.5$ & 24.04$_{-0.21}^{+0.11}$ & -4.52$_{-0.075}^{+0.090}$ & 53 \\
 & 24.33$_{-0.18}^{+0.29}$ & -5.01$_{-0.051}^{+0.057}$ & 67 \\
 & 24.86$_{-0.24}^{+0.23}$ & -5.49$_{-0.079}^{+0.097}$ & 26 \\
 & 25.25$_{-0.16}^{+0.32}$ & -5.58$_{-0.10}^{+0.14}$ & 18 \\
 & 25.85$_{-0.28}^{+0.19}$ & -5.66$_{-0.16}^{+0.26}$ & 13 \\
 & 26.14$_{-0.10}^{+0.37}$ & -6.05$_{-0.17}^{+0.18}$ & 8 \\
$3.5<z<5.5$ & 24.41$_{-0.17}^{+0.16}$ & -5.26$_{-0.10}^{+0.13}$ & 18 \\
 & 24.70$_{-0.14}^{+0.21}$ & -5.90$_{-0.11}^{+0.16}$ & 11 \\
 & 25.12$_{-0.21}^{+0.14}$ & -6.07$_{-0.22}^{+0.25}$ & 5 \\
 & 25.51$_{-0.25}^{+0.10}$ & -5.97$_{-0.20}^{+0.22}$ & 6 \\
 & 25.86$_{-0.25}^{+0.10}$ & -6.36$_{-0.25}^{+0.28}$ & 4 \\
\hline
\end{tabular}
\label{tab:lumfun_vmax}
\end{center}
\end{table*}

\begin{figure*}
\begin{center}
\includegraphics[bb= 20 0 550 660, scale=0.9]{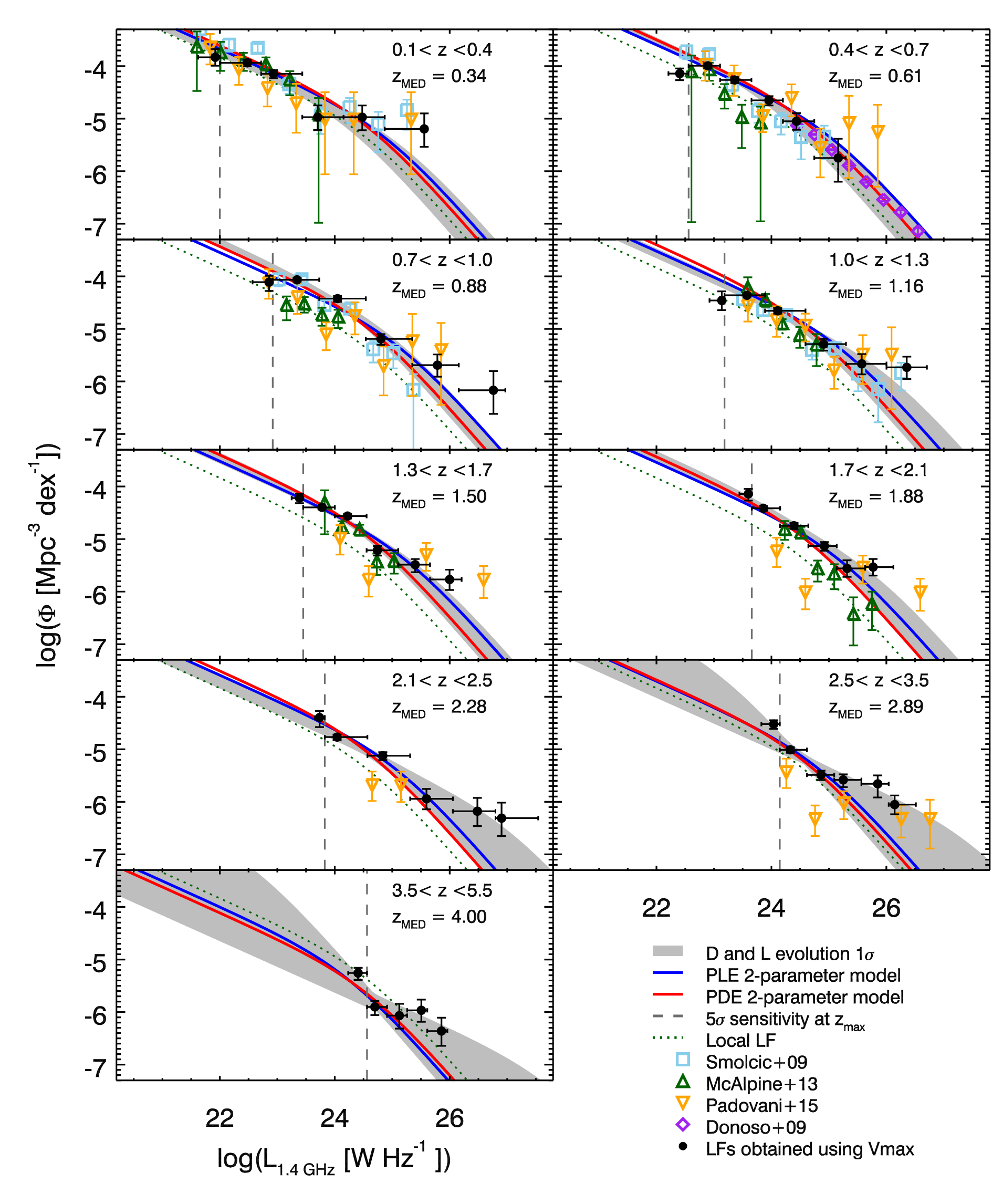}
\caption{Radio AGN 1.4 GHz rest-frame luminosity functions in nine redshift bins out to $z\sim5$ for radio-excess AGN in the COSMOS field.
  The volume densities derived from our radio sample with the $V_\text{max}$ method are shown by the filled black dots.  A $1\sigma$ confidence interval of  the combined luminosity and density evolution in individual redshift bins is shown as the gray shaded area. Pure luminosity evolution and pure density evolution 2-parameter models are shown with blue and red lines, respectively (see text for details).
  The dotted curve in each panel indicates the local luminosity function (Eq.~\ref{eq:locallf}). The vertical line  in each panel shows the $5\sigma$ sensitivity limit (assuming a spectral index of $\alpha=-0.7$) at the high-redshift end of each redshift bin.  
  Various results from the literature, as indicated in the legend, are also shown (each scaled to the median redshift in each bin using the evolution model as reported in the corresponding study).
  \label{fig:lfs}}
\end{center}
\end{figure*}

\subsection{Evolution of the radio AGN luminosity function}
\label{sec:lfevolv}

The cosmic evolution of an astrophysical population is usually parameterized
by  density and luminosity evolution of its local
luminosity function as

\begin{equation}
\label{eq:lfevolv}
\Phi(L,z) = (1+z)^{\alpha_D} \times \Phi_{0}\left[ \frac{L}{(1+z)^{\alpha_L}}
\right],
\end{equation}
where $\alpha_D $ and $\alpha_L$ are the characteristic density and
luminosity evolution parameters, respectively, $L$ is the luminosity,
$\Phi(L,z)$ is the luminosity function at redshift $z$, and
$\Phi_{0}$ is the local luminosity function. 
The  analytic form for the local radio AGN LF, adopted here, and also shown in \f{fig:lfs} , is taken from  \citet{mauch07} and parametrized with two power laws

\begin{equation}
\label{eq:locallf}
\Phi_0(L) = \dfrac{\Phi^*}{(L^*/L)^\alpha + (L^*/L)^\beta},
\end{equation}
where the parameters are the normalization $\Phi^* =
\frac{1}{0.4}10^{-5.5}$~Mpc$^{-3}$~dex$^{-1}$ (scaled to the base of $d \log L$), the knee position  $L^*=10^{24.59}$~W~Hz$^{-1}$, and the bright and faint end slopes $\alpha =-1.27$, and $\beta=-0.49$, respectively. \cite{mauch07} derived their AGN LF using 2661 detections in the 6dFGS–NVSS field with a median redshift of $\text{med}(z)=0.073$ and a span of six decades in luminosities. With such a sample they were able to constrain well both the faint and bright end of the local AGN LF.
For consistency with other studies in the literature and to provide a broad overview, we here take the LF of the radio AGN population as derived by \cite{mauch07}, and model its evolution as it is usually done (using \eq{eq:lfevolv} , see also below). Also, as discussed in \s{sec:intro} \ a two-population model may be more appropriate for modeling the evolution of radio AGN. This is discussed further in \s{sec:unknowns} .

In \f{fig:evolfit} \ we show the best-fit pure density
(PDE; $\alpha_L = 0$) and pure luminosity (PLE; $\alpha_D = 0$)
evolutions for each redshift bin (also tabulated in Tab.~\ref{tab:lumfun_fit}), which can be considered as the two extreme cases of evolution. 
For a conservative approach the outlying, lowest luminosity bins at $z<1.3$  were ignored in the fitting process.

To fit a simple, continuous model to the data, we follow \citet{novak17}, and add a redshift dependent term to the $\alpha_L$, and $\alpha_D$ parameters in \eq{eq:lfevolv} , and model the evolution of the local luminosity function using the following form:

\begin{equation}
\label{eq:lfevolvcont}
\Phi(L,z, \alpha_L, \beta_L, \alpha_D, \beta_D) =  (1+z)^{\alpha_D+z\cdot\beta_D}\times\Phi _{0} \left[ \frac{L}{(1+z)^{\alpha_L+z\cdot\beta_L}}\right],
\end{equation}
where $\alpha_L$, $\beta_L$, $\alpha_D$, and $\beta_D$ are the four free parameters.  For pure luminosity evolution ($\alpha_D=\beta_D=0$) the $\chi^2$ minimization procedure yields best-fit parameter values of $\alpha_L=2.88\pm0.82$, $\beta_L=-0.84\pm0.34$, while for pure density evolution  ($\alpha_L=\beta_L=0$) the best-fit parameters are $\alpha_D=2.00\pm0.18$, $\beta_D=-0.60\pm0.14$. 
Fitting for all four parameters simultaneously yields a strong degeneracy between the parameters and mainly differs from the two two-parameter fits at the high-luminosity end in  high-redshift bins ($z>2.1$). As discussed in more detail in \s{sec:unknowns} , the volume densities in these particular bins are the most sensitive to the assumption of a simple synchrotron power law for the K correction to rest-frame 1.4~GHz luminosity from the observed 3 GHz flux density. 
Hence, as  the typical AGN spectrum at radio frequencies in deep surveys, such as the VLA-COSMOS 3 GHz survey, to date is not constrained, and further, given the degeneracies inherent to the four-parameter fit, 
we hereafter adopt the simple continuous two-parameter models as a representation of the evolution of the VLA-COSMOS 3 GHz (radio-excess) AGN. Also, simple, pure luminosity (density) evolution models are commonly used in the literature \citep[e.g.,][]{sadler07,smo09a,donoso09,mcalpine13,padovani15}.

\begin{figure}
\begin{center}
\includegraphics[bb=0 0 550 339, width=\columnwidth]{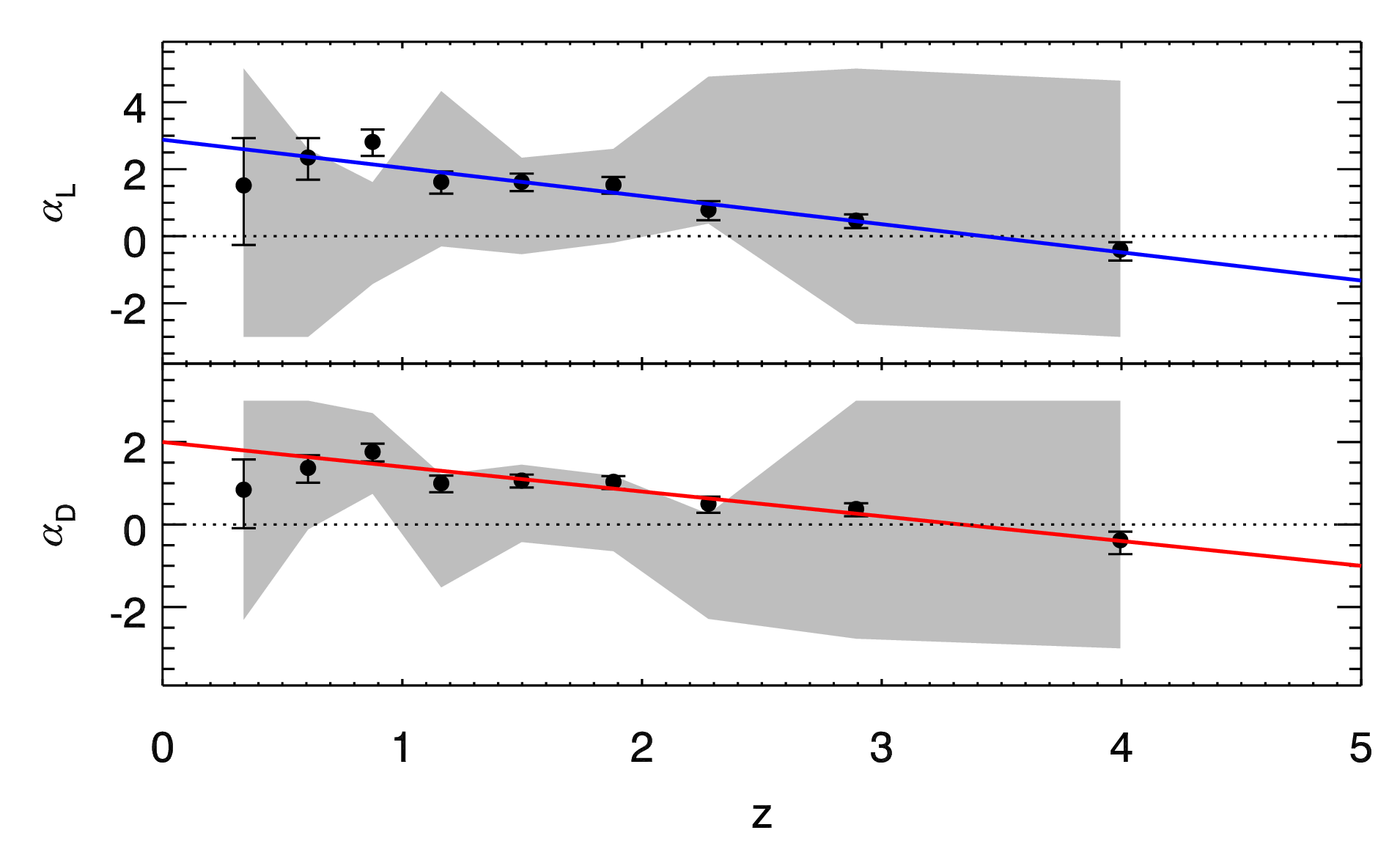}
\caption{ Best-fit parameters for the evolution of the local LF with
redshift. The best-fit pure luminosity (density) evolution in a given redshift bin is shown by the black points in the top (bottom) panel. The blue (red) line in the top (bottom) panel shows the simple, continuous 2-parameter pure luminosity (density) evolution model (see text for details). Gray shaded areas in both panels correspond to the 68\% confidence interval for a combined luminosity and density evolution. The large uncertainty in the combined fit is due to parameter degeneracy. 
\label{fig:evolfit}}
\end{center}
\end{figure}

\begin{figure}
\begin{center}
\includegraphics[bb=25 0 550 339, scale=0.48]{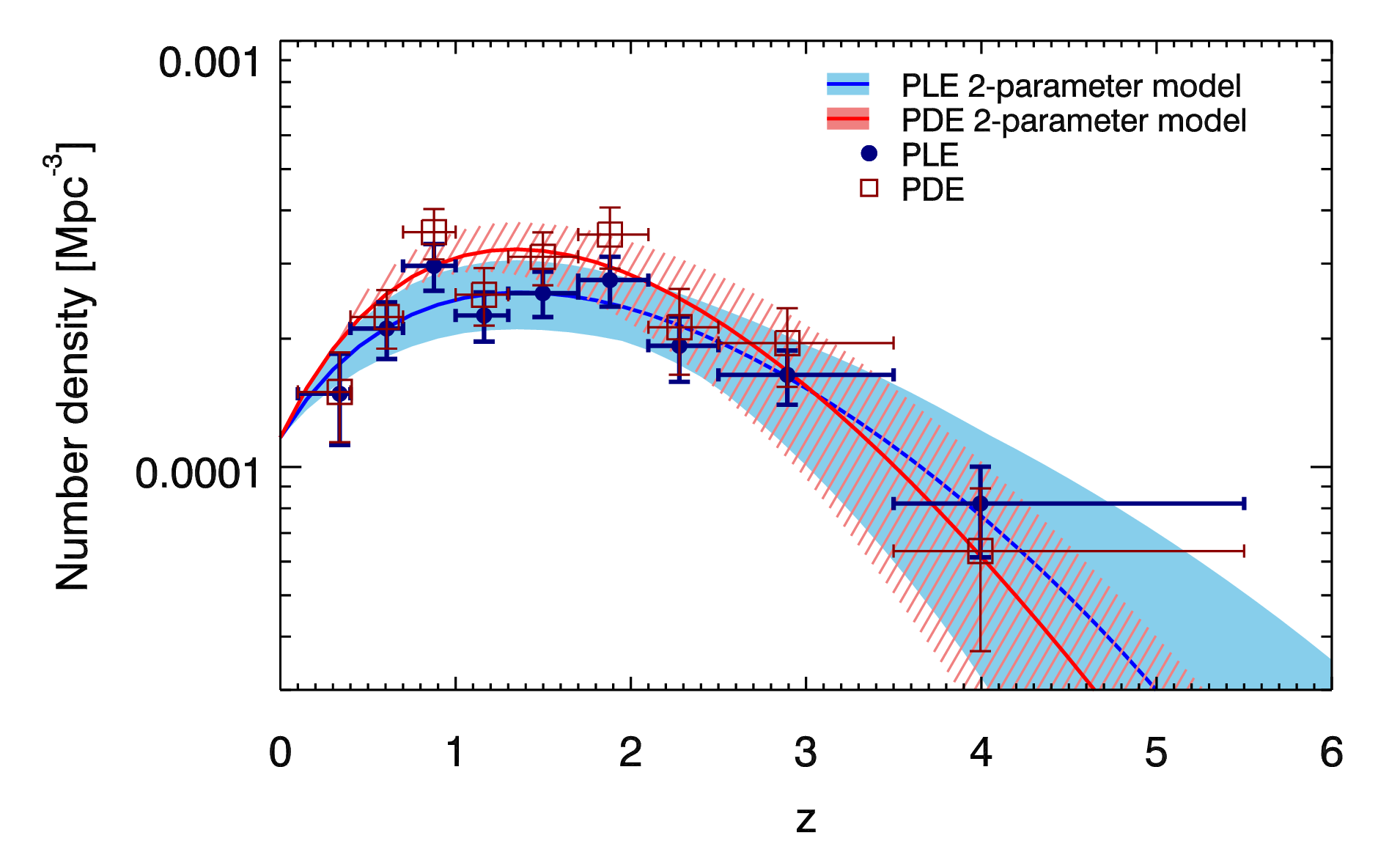}
\includegraphics[bb=0 0 550 339, scale=0.45]{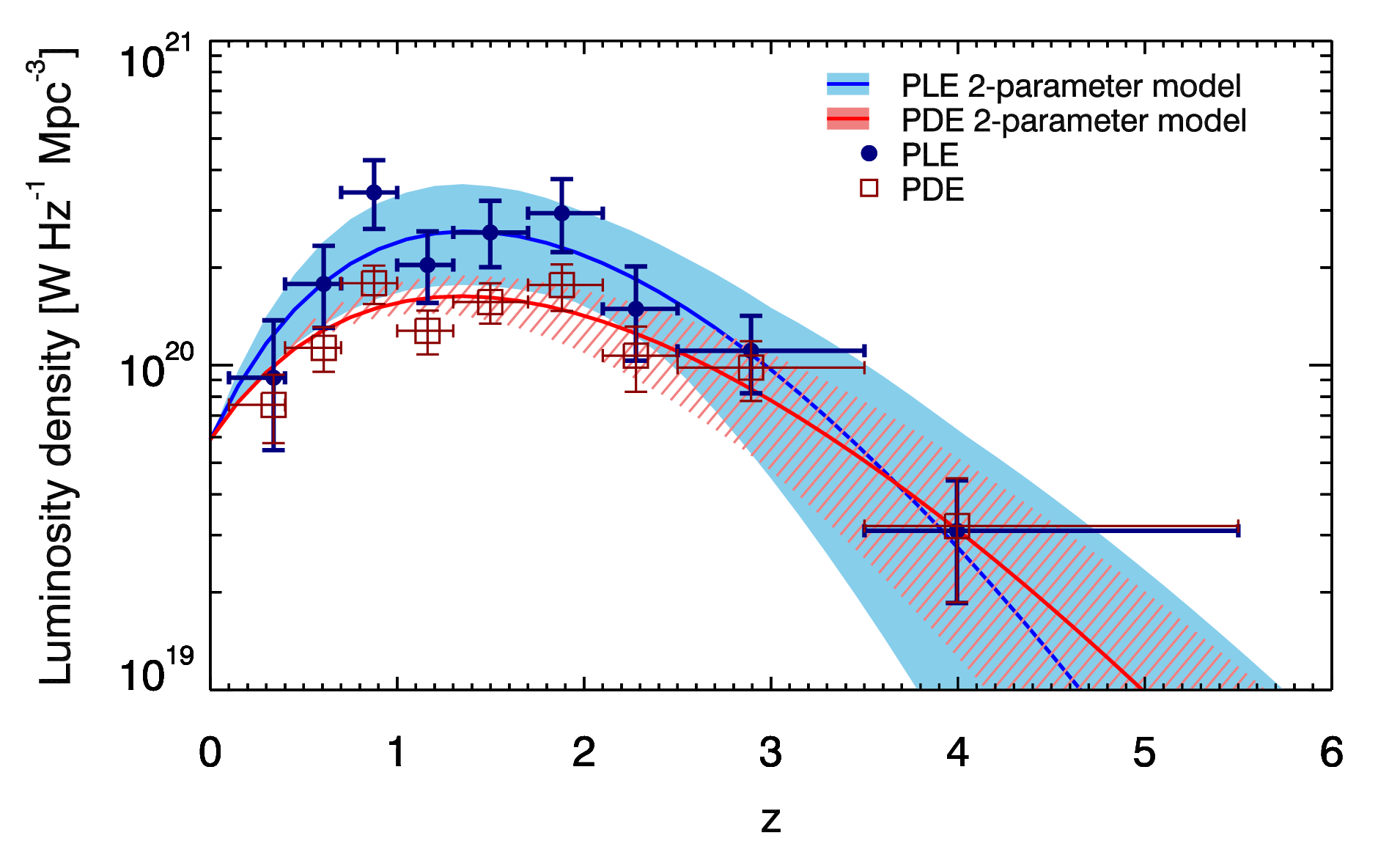}
\caption{ Redshift evolution of the number density (top panel), and luminosity density (bottom panel) for our radio AGN (for models as indicated in the panels; see text for details). The number density was integrated with a lower boundary of \lum ~$=10^{22}$~\wh . Decreasing the boundary would systematically shift the number densities to higher values (for example, by a factor of 3-5 if \lum ~$=10^{21}$~\wh \ were used). The luminosity density integral has a much weaker dependence on the integration limits (if it is broad enough) as the bulk of the integral is constrained close to ($L^*,\Phi^*$).
  \label{fig:ld}}
\end{center}
\end{figure}

\subsection{Cosmic evolution of the number and radio luminosity densities }
\label{sec:ld}

In \f{fig:ld} \ we show the redshift evolution of the number and luminosity densities using our two-parameter evolution models and the best-fit pure luminosity and density evolutions in each redshift bin. The number density in a given redshift bin was obtained by integrating the corresponding luminosity function over the logarithm of luminosity, $\int \Phi (L_\mathrm{1.4GHz}) \, d(\log{L_\mathrm{1.4GHz}})$, and the luminosity density by integrating the product of luminosity and volume density over   the logarithm of luminosity, i.e., $\int L_\mathrm{1.4GHz} \times \Phi  (L_\mathrm{1.4GHz}) \, d(\log{L_\mathrm{1.4GHz}})$. 

The number density of our radio AGN increases by a factor of $\sim2-3$ from $z=0$ to $z\sim1.5$, beyond which it decreases, reaching a number density equivalent to that in the local universe at $z\sim3.5$ and further decreasing beyond this redshift. The luminosity density shows a similar behavior. It rises by a factor of $\sim2-4$ from $z=0$ to $z\sim1.5$, beyond which it decreases, at $z\sim5$ reaching a value about an order of magnitude lower than the luminosity density derived at $z=0$.

\begin{table}
\begin{center}
\caption{Best-fit LF evolution parameters in individual redshift bins.}
\renewcommand{\arraystretch}{1.1}
\begin{tabular}[t]{c c c}
\hline
Redshift & PDE & PLE \\
Med $(z)$ & $\alpha_D$ & $\alpha_L$\\
\hline
0.338 & 0.844 $\pm$ 0.83 & 1.52 $\pm$ 1.6 \\
0.607 & 1.37 $\pm$ 0.33 & 2.35 $\pm$ 0.62 \\
0.876 & 1.76 $\pm$ 0.22 & 2.81 $\pm$ 0.39 \\
1.16 & 1.00 $\pm$ 0.20 & 1.62 $\pm$ 0.33 \\
1.50 & 1.07 $\pm$ 0.16 & 1.62 $\pm$ 0.26 \\
1.88 & 1.04 $\pm$ 0.16 & 1.53 $\pm$ 0.25 \\
2.28 & 0.502 $\pm$ 0.20 & 0.788 $\pm$ 0.28 \\
2.89 & 0.375 $\pm$ 0.16 & 0.467 $\pm$ 0.20 \\
4.00 & -0.381 $\pm$ 0.27 & -0.405 $\pm$ 0.27 \\
\hline
\end{tabular}
\label{tab:lumfun_fit}
\end{center}
\end{table}

\subsection{Comparison with the literature}
\label{sec:compare}

Based on the 2SLAQ Luminous Red Galaxy Survey containing $\sim400$ galaxies in a volume-limited sample at $0.4<z<0.7,$ \citet{sadler07} have found that their radio AGN (\lum ~$\approx 10^{24}-10^{27}$~\wh ) undergo significant evolution since $z\sim0.7$, parametrized with a pure luminosity evolution parameter $\alpha_L=2.0\pm0.3$. \citet{donoso09} found fully consistent results, but with considerably smaller error bars as they used a sample of over 14,000 radio AGN. These derivations are in very good agreement with the luminosity function derived here (see $0.4<z<0.7$ bin in  \f{fig:lfs} ) and with the pure luminosity evolution we find for the VLA-COSMOS radio-excess AGN, detected at 3 GHz (see top panel of \f{fig:evolfit} ).

\citet{mcalpine13} studied the evolution of faint radio AGN in the VIDEO-{\em XMM3} field out to $z\sim2.5$. By fitting a combined evolution of star-forming and AGN galaxies, they found a slightly weaker evolution than that inferred by \citet[][see Table 4 in \citealt{mcalpine13}]{sadler07}. The authors argue that this is because the evolution of the VIDEO-{\em XMM3} radio AGN is primarily driven by the higher redshift range ($0.9<z<2.5$), not constrained by the 2SLAQ survey. The pure luminosity evolution inferred by \citet[][$\alpha_L=1.2\pm0.2$]{mcalpine13} is consistent with the average $\alpha_L$ derived here in the equivalent ($0.9<z<2.5$) redshift range (for example, for $z=1.75$, $\alpha_L+z\cdot\beta_L=1.4\pm1.0$ for our pure luminosity two-parameter fit; see top panel of \f{fig:evolfit} ). In each redshift range the VIDEO-{\em XMM3}-based volume densitites of their AGN (shown in \f{fig:lfs} ), which were identified during the photometric-redshift estimation process via template fitting (see their Sec. 6.2.1.), are slightly below those derived here, particularly at the high-luminosity end in each redshift bin (see \f{fig:lfs} ). This underestimate of the number density of AGN at the high-luminosity end, and especially at $z>1$, however, has already been observed and reported by \citet[][see their Fig.~8]{mcalpine13}.

Based on the VLA-COSMOS 1.4~GHz survey \citet{smo09a} have derived luminosity functions for their rest-frame color selected AGN out to $z=1.3$, which is also shown in \f{fig:lfs} . Overall, there is good agreement between the derivations based on the (shallower) 1.4 GHz survey, and the 3 GHz survey used here. 
\citet{smo09a} inferred a pure luminosity (density) evolution out to $z=1.3$ for their AGN of $\alpha_L=0.8\pm0.1$ ($\alpha_D=1.1\pm0.1$). While their pure density evolution is in agreement with that derived here, the pure luminosity evolution is here derived to be slightly higher; i.e., $\alpha_L(z=1.3)= 1.6\pm0.2$, and $\alpha_L+z\cdot\beta_L=1.8\pm0.9$ for $z=1.3$, and our pure luminosity two-parameter fit. The differences are due to a combination of factors, such as  i) the $\sim3$ times increased sensitivity of the VLA-COSMOS 3 GHz, compared to the 1.4 GHz Large projects,  ii) the different local luminosity functions used in the two studies; while the \citealt{mauch07} local luminosity function is used here, the \citealt{sadler02} local luminosity function was used by \citealt{smo09a}, and iii) differently selected AGN. \citet{smo09a} selected the radio AGN by requiring host galaxies with red rest-frame colors \citep{smo08} that mimic the spectroscopically based identification \citep{bpt} commonly used in the local universe, while we use the radio-excess criterion to identify our AGN, which includes both host galaxies with red rest-frame colors, but also AGN selected via X-ray, and mid-infrared (MIR) criteria within blue host galaxies (see \s{sec:unknowns} , and \f{fig:subpop} \ below; see also \citealt{delvecchio17}).

\citet{bonzini13} identified a sample of ($z\leq4$) radio-loud AGN in the E-CDFS field by requiring that at a given redshift, they lie below the $2\sigma$ deviation from the average observed $24~\mu$m-to-1.4~GHz flux ratio obtained using the M82 galaxy template. This criterion is similar to that used here to select radio-excess AGN (see Sec. 4.3.2. in \citealt{delvecchio17}), and thus in \f{fig:lfs} \ we also compare the luminosity functions derived here with those derived by \citet{padovani15} for the radio-loud AGN in the E-CDFS field. Overall, the E-CDFS-based luminosity functions agree well with those derived here out to $z=1.3$. Beyond this redshift a higher degree of discrepancy is observed. This is likely related to small number statistics given the size of the E-CDFS survey (0.3 square degrees) resulting in most of the E-CDFS volume density values at $z>1.3$ constrained by bins containing fewer than three sources (cf.\ Table 5 in \citealt{padovani15} and \t{tab:lumfun_vmax} \ here).

\section{ Radio-mode feedback considerations  } 
\label{sec:feedback}

In this Section we derive the cosmic evolution of the radio AGN kinetic luminosity density (\s{sec:lkin} ), and compare this value with the SAGE semi-analytic model (\s{sec:sage} ).

\subsection{Cosmic evolution of the radio AGN kinetic luminosity density}\label{sec:lkin}

The AGN studied here have, by construction, an observed  excess of radio emission relative to that expected from star formation processes within their host galaxies. Thus, the origin of the observed 3~GHz radio emission in these sources is related to the AGN itself, i.e.,\ to the energy released through mass accretion onto the central SMBH. This energy can  be efficiently radiated away and/or channeled in kinetic form via collimated jets expanding through and beyond the host galaxy and is observable at radio wavelengths. 
The observed  radio emission  in our sources (albeit mostly unresolved or barely resolved at $0.75\arcsec$ resolution) can thus be predominantly attributed to the synchrotron emission of relativistic particles within the jet structures of the sources (i.e., core, jets, and lobes). However, the observed radio emission amounts to only a fraction of the total kinetic luminosity\footnote{The total kinetic luminosity is taken to be equal the total energy transported by the jets during the lifetime of the radio source \citep{willott99}.} of the jets, while a much larger fraction is stored as the internal energy of the jets and lobes and is lost to the environment via the work carried out by the expanding radio jets \citep[e.g.,][]{willott99}. The latter is of particular interest as the energy (per unit time) deposited into the surroundings and dissipated can be directly linked to the heating of the surrounding medium generated by the jets of such AGN (see \citealt{mcnamara07,mcnamara12}  for reviews). 

A simple scaling relation between the monochromatic radio and kinetic luminosities has been long sought after. In  Appendix~\ref{sec:scl} we give a detailed overview of such scaling relations, commonly used in the literature \citep{willott99, birzan04, birzan08, merloni07, cavagnolo10, osullivan11, daly12, godfrey16}. We hereafter adopt the relation derived by \citet{willott99} at 151 MHz rest-frame frequency and converted to 1.4~GHz rest-frame luminosity \citep{heckman14}, given as 
\begin{equation}
\label{eq:willott}
\log{L_\mathrm{kin} (L_\mathrm{L_1.4GHz}) }= 0.86\cdot\log{L_\mathrm{L_1.4GHz}} + 14.08 + 1.5\log{f_\mathrm{W}}$$, 
\end{equation}
where $L_\mathrm{kin}$ is the kinetic luminosity in units of W, $L_\mathrm{L_1.4GHz}$ is the 1.4~GHz rest-frame luminosity in units of W/Hz, and $f_W$ is an uncertainty parameter (see below). 

\citet{willott99} estimated the kinetic energy using minimum energy arguments, i.e.,\ computing the minimum  energy stored in the lobes to produce the observed synchrotron luminosity, considering the age of the source and efficiency of conversion of the kinetic luminosity into the internal energy of the observed synchrotron emission. 
These authors have folded all uncertainty factors in the calculation, such as departure from minimum energy conditions, uncertainty in the energy in nonradiating particles, and the composition of the jet, into one parameter, $f_\mathrm{W}$, estimated to lie in the range of $f_\mathrm{W}\approx1-20$. For $f_\mathrm{W}=15,$ the normalization is close to that of the jet kinetic luminosities computed through X-ray observations of galaxy clusters with cavities induced in the hot, X-ray emitting instracluster gas  by the radio jets and lobes \citep{birzan04, birzan08, merloni07, cavagnolo10, osullivan11}, and for $f_\mathrm{W}=4$ the normalization matches the relation derived for powerful radio galaxies using equations of strong shock physics \citep[see][and references therein]{daly12}. We, however, also consider the full range of the uncertainty parameter ($f_W=1-20$; see below). 

The kinetic luminosity density at a given redshift is computed as the integral of the kinetic luminosity density over (the logarithm of the) monochromatic 1.4~GHz radio luminosity,
\begin{equation}
\label{eq:lkin}
\Omega_\mathrm{kin}(z)=\int L_\mathrm{kin}  (L_\mathrm{L_1.4GHz}) \times \Phi  (L_\mathrm{L_1.4GHz}) \, d(\log{L_\mathrm{1.4GHz})}.
\end{equation}
In \f{fig:lkinevolv} \ we show the cosmic evolution of the kinetic luminosity density derived using our two-parameter continuous evolution model (Eq.~\ref{eq:lfevolvcont}) and the scaling relation between monochromatic radio luminosity and kinetic luminosity of \citet[][Eq.~\ref{eq:willott}]{willott99} with uncertainty parameters $f_W=4$ and $15$, and the range encompassed by the extreme values of $f_W$ ($=1, 20$).  
The various $f_W$ values only systematically shift the derived volume-averaged kinetic luminosity density as a function of redshift. Furthermore, the derivation for $f_W=1$ can be considered as a robust lower limit, as any other scaling relation available in the literature would have resulted in systematically higher values (see Appendix~\ref{sec:scl} for details). Under the assumptions made, the kinetic luminosity density rises by about a factor of three from $z=0$ to $z\sim1.5$ and decreases thereafter by close to two orders of magnitude by $z\sim5$. This holds for both of our (pure luminosity and pure density) two-parameter models, as also illustrated in \f{fig:lkinevolv} .

\begin{figure}
\begin{center}
\includegraphics[bb=0 0 550 339, width=\columnwidth]{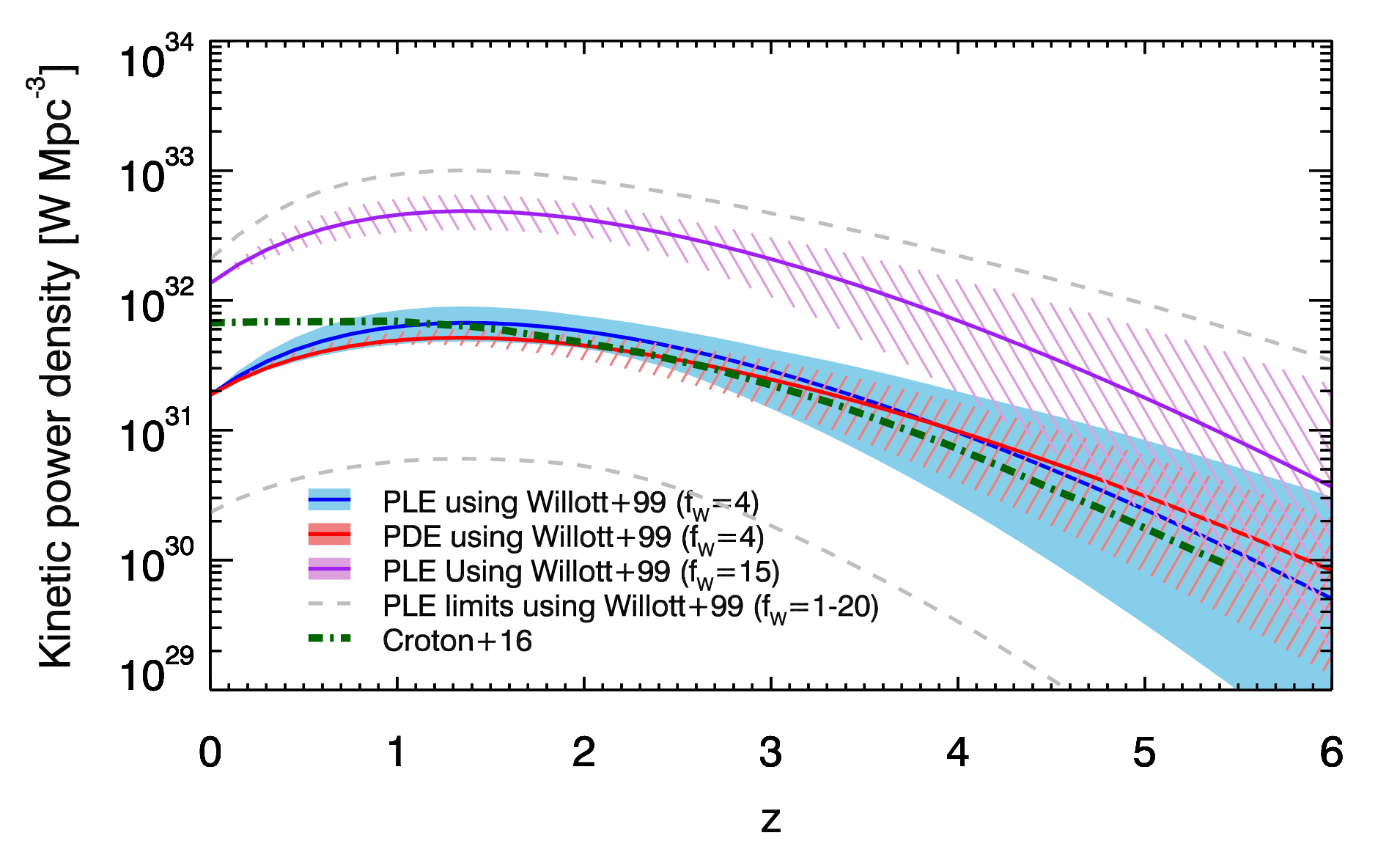}
\caption{  Cosmic evolution of the kinetic luminosity density for our radio AGN. The kinetic luminosity density was derived using Eq.~\ref{eq:lkin}, and the \citet{willott99} relation between monochromatic (1.4~GHz rest-frame) radio luminosity and kinetic luminosity (Eq.~\ref{eq:willott}) with $f_W=4$ (pure luminosity evolution, PLE, model: blue full curve  and blue shaded area showing the $1\sigma$ uncertainty), $f_W=15$ (purple full curve, and purple-hatched area showing the $1\sigma$ uncertainty), and $f_W=1$, and $20$, with folded $1\sigma$ fitting errors (lower and upper gray dashed curves, respectively). Also shown is the kinetic luminosity density assumed in the SAGE semi-analytic cosmological model (green dash-dotted curve; \citealt{croton16}). For comparison, the evolution of the kinetic luminosity density for our radio AGN for the pure density evolution, PDE, model using only $f_W=4$ (Eq.~\ref{eq:scl}) is also shown with the red full curve and red hatched area. To avoid overcrowding the panel, the PDE result for $f_W=15$ is not shown, but it would be only systematically higher compared to the $f_W=4$ PDE model result and coincident with that for PLE with the corresponding $f_W=15$ value. See also \f{fig:lkinapp} \ for results based on other scaling relations.
  \label{fig:lkinevolv}}
\end{center}
\end{figure}

\subsection{Comparison with the SAGE semi-analytic  model}
\label{sec:sage}

In \f{fig:lkinevolv} \ we compare the redshift evolution of the COSMOS radio AGN kinetic luminosity density with that from the SAGE model \citep{croton16}. The SAGE model is a significant update of the semi-analytic model by \citet{croton06}, including implementation of a more complex cycle between gas cooling and the radio-mode AGN heating. 
It is assumed that hot gas accretes onto the central black hole
following the Bondi-Hoyle accretion  \citep{bondi52}, but including a, so-called, radio-mode efficiency parameter ($\kappa_R$; see Eq.~16 in \citealt{croton16}). This  parameter is used to modulate the strength of black hole accretion ($\dot{m}$), assumed to be related to the luminosity of the black hole in radio mode in the standard way, as $L = \eta \dot{m} c^2$, where $\eta=0.1$ is the standard efficiency of mass to energy conversion, and $c$ is the speed of light. In the SAGE model this luminosity is taken as the source of heating that offsets the energy losses of the cooling gas, such that the heating rate is simply the ratio between the luminosity and the specific (per unit mass) energy of gas in the hot halo (see Eq.~17 and Sec.~9.1. in \citealt{croton16}). The accretion luminosity assumed in the SAGE model can be taken as an equivalent to the, here derived, jet kinetic luminosity assuming that the bulk of the accretion energy is channeled in kinetic (rather than radiative) form. Hence, in \f{fig:lkinevolv} \ we can compare the redshift evolution of the COSMOS radio AGN kinetic luminosity density with the volume-averaged accretion luminosity  responsible for radio-mode feedback in the SAGE model, which is equivalent to that  obtained by \citet[][their Fig.~3]{croton06}, but scaled by the radio-mode efficiency parameter ($\kappa_R=0.08$; see Sec.~9.1.1. in \citealt{croton16}).

As seen in \f{fig:lkinevolv} \ at $z\gtrsim1,$ we find a remarkably similar slope of the redshift evolution of the COSMOS radio AGN kinetic luminosity density and the radio-mode accretion luminosity used in the SAGE model with the absolute values on the same order for $f_W=4$ (the uncertainty parameter in the radio luminosity -- kinetic luminosity relation; \eq{eq:willott} ). At $z<1$ we find a steeper evolution than that in the SAGE model, for which our $z=0$ value is a factor of 3-4 lower ($f_W=4$). For $f_W=15$ the kinetic luminosity density is at every redshift systematically higher than the radio-mode accretion luminosity in the model, while for the robust lower limit value ($f_W=1$) the observationally based  kinetic luminosity density is at each redshift about an order of magnitude lower than that in the SAGE model. Overall, for the most likely values of the normalization of the  radio luminosity -- kinetic luminosity relation ($f_W=4,15$; see previous section, and Appendix~\ref{sec:scl}), the redshift evolution of the COSMOS radio AGN kinetic luminosity density is either on the same scale as that in the model or  is systematically higher by about an order of magnitude or even higher if other scaling relations are used (see \f{fig:lkinapp} ). This would suggest that, in both cases, the energy deposited by faint radio AGN into their environment may be sufficient to offset the cooling energy losses, as postulated in the model. However, there is still a non-negligible number of simplifications and unknowns inherent in both the semi-analytic models, and the observational results, as discussed in the next section.

\section{Discussing the unknowns}
\label{sec:unknowns}

Observational studies of radio-selected AGN find two radio-luminous AGN populations with distinct host galaxy and AGN properties \citep{smo09, smo09a, smo15, smo16, hardcastle07, buttiglione10, heckman14, bonzini15, padovani15, padovani16, tadhunter16}. One population is consistent with the standard, unified model AGN picture, in which the accretion occurs in a radiatively efficient manner at high Eddington rates ($1-10\%\lesssim\lambda_\mathrm{Edd}\lesssim 100\%$). Evidence has been presented that this class is fueled by the cold intragalactic medium (IGM) phase and that it is not too likely to launch collimated jets (observable at radio wavelengths). The second population, however, exhibits radiatively inefficient accretion related to low Eddington ratios ($\lambda_\mathrm{Edd}\lesssim 1-10\%$) and may be fueled by the hot phase of the IGM. It has been shown that this population is highly efficient in collimated jet production. The observed difference in Eddington ratio between the two populations can be linked to the switch between the standard accretion flow model, i.e., radiatively efficient, geometrically thin (but optically thick) disk accretion flow \citep{shakura73}, and a radiatively inefficient,  geometrically thick (but optically thin)  accretion flow \citep{esin97, narayan98}, occurring at accretion rates below a certain Eddington ratio (1-10\%; \citealt{rees82, narayan94, meier02, fanidakis11}). Further differences have been found demonstrating that the population of radiatively efficient accretors dominates the radio-AGN number densitites at the bright end (e.g., \lum~$\gtrsim10^{26}$~\wh for $z<0.3$; \citealt{pracy16}) and that this population evolves more rapidly with cosmic time \citep[e.g.,][]{willott01,pracy16}. 

The radio AGN studied here have been selected based on an excess of their radio emission relative to that expected from the (IR-based) star formation rates in their host galaxies, thus assuring that $\gtrsim80\%$ of the radio emission arises from the AGN core, jet, and/or lobe component. In \f{fig:subpop} \ we plot the absolute and fractional contributions of X-ray-,  MIR-selected AGN, and the remaining AGN (not identified via X-ray or MIR emission), those hosted by red, quiescent galaxies, and those hosted by galaxies with blue or green rest-frame colors, implying substantial star formation activity in the hosts (see \citealt{smo17b} for details). The X-ray and MIR regimes provide an efficient approach to identify radiatively efficient AGN, while red, quiescent host galaxies of radio AGN are shown to contain AGN with systematically lower radiative AGN luminosities (see, e.g., Fig.~7 in \citealt{smo17b}); further, this selection can be used to trace radiatively inefficient AGN (at least in the local universe; e.g., \citealt{smo09}). 

As seen in \f{fig:subpop} , at $z\lesssim1$ our radio-excess AGN are composed of similar fractions of i) red, quiescent galaxies and ii) X-ray, MIR, and those hosted by star-forming galaxies, but not identified via X-ray or MIR emission. Beyond $z=1$ the fraction of red, quiescent galaxies decreases steeply to a minimal fraction by $z\sim2$. Hence, we can conclude that AGN with observational signatures of radiatively efficient accretion flows comprise a non-negligible fraction of our radio-excess AGN (30-40\% X-ray- and MIR-selected AGN). If only a fraction of the radio-excess AGN hosted by blue or green, star-forming galaxies also contained radiatively efficient AGN, which could be the case given the expected cold gas supply in such, actively star-forming galaxies \citep[e.g.,][]{vito14}, the fraction of radiatively efficient AGN in our radio-excess sample would rise even further. This implies that our radio-excess AGN are likely a mix of radiatively efficient and inefficient black hole accretion flows, shown in other studies to evolve differently with cosmic time \citep[e.g.,][]{willott01, best14, pracy16}. 

The volume densities derived here for our radio AGN are, at the high-luminosity end in every redshift range, somewhat higher than our best-fit  pure luminosity/density two-parameter models (see \f{fig:lfs} ). This could potentially be attributed to the rise given the faster evolution of the high-luminosity radio AGN that are consistent with the radiatively efficient accretion flow population. However,  these particular, high-luminosity bins at each redshift are the most sensitive to the assumption of a simple synchrotron power law for the conversion  from the observed 3 GHz flux density to rest-frame 1.4~GHz luminosity.
For example, assuming a spectral index of $\alpha=-0.7$ for all our AGN instead of using the observed value, if available, which steepens toward high redshift (see \f{fig:alphas} ), this would have no effect on the low-luminosity volume densities in the high-redshift bins, but the high-luminosity bin values would decrease to be consistent with the pure luminosity (density) two-parameter models. 

If the typical radio AGN spectrum were to steepen toward the high-frequency end owing to synchrotron energy losses (e.g., \citealt{miley80}), as sampled by the observed frequency of 3 GHz at $z>2$ (corresponding to rest-frame frequencies of $>9$~GHz) then, for example, a broken power law, rather than a simple, single power-law assumption would be more appropriate for the K correction. On the other hand, in this case, the simple, single power-law assumption of $\alpha=-0.7$ could result in a more correct 1.4 GHz rest-frame luminosity value than the directly observed higher spectral index values. 
     
Although our radio-excess AGN are likely a mix of radiatively efficient and inefficient black hole accretion flows
it could be assumed that, if heating from radio outflows  is indeed occurring in these systems, this heating operates in a similar fashion in the various AGN within our sample, as postulated in the SAGE model. In the SAGE model it is taken that cooling flows develop in any halo with mass $\gtrsim2.5\cdot10^{11}$~\msol , and further, that Bondi-Hoyle accretion of the hot (subdominant) component of the cooling gas, which fills the space between the (dominant) cold cloud component, is responsible for the luminosity of the SMBH in radio mode. 
Taking the stellar mass -- halo mass relation of \citet{behroozi10}, the halo mass threshold beyond which radio-mode feedback occurs can be converted to a stellar mass of $\gtrsim5\cdot10^{9}$~\msol . The stellar masses of our radio AGN satisfy this criterion (see \f{fig:mstar} ), and can, hence, be taken as a representative population of that assumed in the SAGE model to be responsible for radio-mode feedback. In that respect, the similarity between the here derived redshift evolution of the kinetic luminosity density and that assumed in the SAGE semi-analytic model (at least at $z>1$) is astonishing. As a further step a more complex treatment of radio AGN in semi-analytic models, taking into account the two dominant, thin- and thick-disk, accretion flows and the spins of SMBHs, would be desirable \citep[e.g.,][]{fanidakis11}. However, from an observational perspective
likely the largest source of uncertainty in studies of radio-mode feedback 
is the uncertainty of the monochromatic radio luminosity to kinetic luminosity conversion. The relation used here contains an uncertainty factor on the order of 20 ($f_W$ in \eq{eq:willott} ; \citealt{willott99}) and various relations commonly used in the literature are summarized in Appendix~\ref{sec:scl}. The kinetic luminosity has been demonstrated to be not only a function of monochromatic radio luminosity, but also of other parameters such as the synchrotron electron age of the AGN and its environment \citep{birzan08, hardcastle14, kapinska15}. A better constraint of the relation as a function of redshift and other physical parameters is still awaited. In this context, deep radio surveys, observed in a large enough range of radio frequency, allowing us to compute synchrotron ages, could provide an important step toward better understanding of the relevance of radio-mode AGN feedback in massive galaxy formation.
      
      \begin{figure}
\begin{center}
\includegraphics[bb=0 15 432 432, width=\columnwidth]{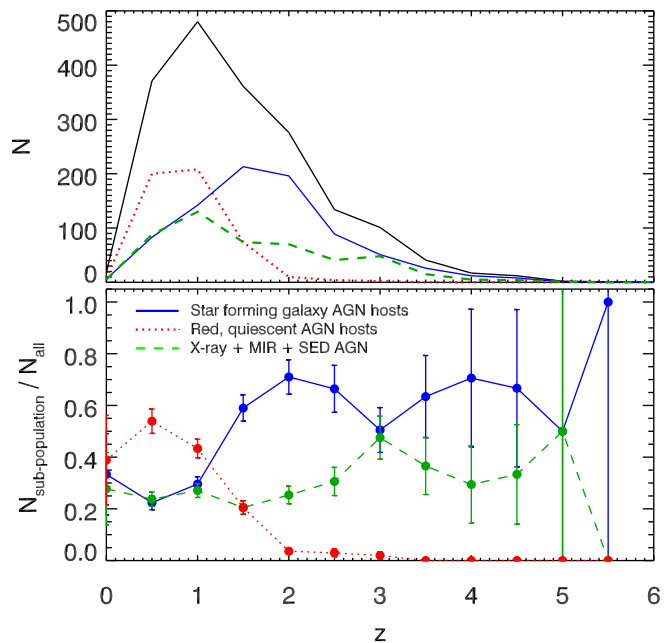}
\caption{ Absolute (top panel) and fractional (bottom panel) contributions of various subpopulations (as indicated in the bottom panel) to the full radio-excess AGN sample (black curve) as a function of redshift. 
  \label{fig:subpop}}
\end{center}
\end{figure}

 \section{Summary and conclusions}     
 \label{sec:summary}
   
We have used a sample of 1,814 radio AGN selected from the VLA-COSMOS 3~GHz Large Project, and with NIR-detected (COSMOS2015 catalog) counterparts with photometric or spectroscopic redshifts. These were selected to have radio emission in $>3\sigma$ excess of that expected from the IR-based star formation rates of their hosts. Such a criterion assures that at least 80\% of their radio emission is due to the AGN component. This sample of AGN reaches out to a redshift of $z < 6$, comprises 1.4 GHz rest-frame radio luminosities in the range of $10^{22}-10^{27}$~\wh , and the typical stellar masses of the AGN are within $\sim3\times(10^{10}-10^{11})$~\msol .

Using the $1/V_\mathrm{max}$ method we derived the luminosity functions of the radio AGN out to $z\sim5$. We further constrained the evolution of this population with continuous models of pure density and pure luminosity  evolutions finding best-fit parameters of $\Phi^*\propto(1+z)^{(2.00\pm0.18)-(0.60\pm0.14)z}$ and $L^*\propto(1+z)^{(2.88\pm0.82)-(0.84\pm0.34)z}$, respectively. We find a turnover in number and luminosity densities of the population at $z\approx1.5$. 

The 1.4\,GHz luminosity was converted to kinetic luminosity via an analytically motivated relation. Taking into account the full range of uncertainty, we derived the cosmic evolution of the kinetic luminosity density provided by our AGN, which we compare to the radio-mode AGN feedback assumed in the SAGE model.
We find that the kinetic luminosity exerted by our radio AGN may be high enough at each cosmic epoch since $z\sim5$ to balance the radiative cooling of the hot gas, as assumed in the model. However, although our findings  support the idea of radio-mode AGN feedback as a cosmologically relevant process in massive galaxy formation, many simplifications in both the observational and semi-analytic approaches still remain and need to be resolved before robust conclusions can be reached.

\section*{Acknowledgement}
This research was funded by the European Union's Seventh Frame-work programs
under grant agreements 333654 (CIG, "AGN feedback"). VS, MN, ID, and LC acknowledge support from the European Union's Seventh
Frame-work program under grant agreement 337595 (ERC Starting Grant, "CoSMass"). MB acknowledges support from the PRIN-INAF 2014. ES acknowledges funding from the European Research Council (ERC) under the European Union's Horizon 2020 research and innovation program (grant agreement n$^\circ$ 694343).

\bibliographystyle{apj} 
\bibliography{refs}

\appendix

\section{Scaling relations between monochromatic radio luminosity and  kinetic luminosity}
\label{sec:scl}

Here we summarize the scaling relations between monochromatic radio luminosity and  kinetic jet luminosity commonly used in the literature and illustrated in 
  \f{fig:pcav} .

A commonly used technique to estimate the kinetic luminosity is via X-ray observations of galaxy clusters hosting radio galaxies, whose 
jets and lobes, which succeeded in propagating beyond their host galaxies and into the intracluster gas, induce cavities in the hot X-ray-emitting gas and inflate buoyantly rising bubbles (see, e.g., \citealt{birzan04,allen06,birzan08,cavagnolo10,osullivan11}). As the total energy within a bubble is the sum of its internal energy and the  work carried out by its inflation, the kinetic luminosity is given by the ratio of this total energy and the age of the bubble.

\citet{birzan08} derived  kinetic luminosities for a sample of 24 radio galaxies (\lum~$<10^{28}$~W/Hz, $z<0.4$) in  clusters or groups and found a rather flat ($A=0.35$) slope of the correlation (see \eq{eq:scl} \ below), albeit with a large scatter of $\sigma\sim0.84$~dex.\footnote{On theoretical grounds, a large scatter is expected as the monochromatic radio luminosity  also depends on the age and size of the source and the ambient medium  the jet is penetrating \citep[i.e., the density gradient along the path of the jet; e.g., ][]{anbaan12,hardcastle14}. } This sample was augmented by \citet{cavagnolo10} and \citet{osullivan11} to better constrain the faint end (\lum~$\sim10^{20}-10^{24}$~\wh), which resulted in a revised steeper ($A\approx0.6-0.9$) slope of the relation with a scatter of $\sigma\sim0.7$~dex.  Using similar samples of radio AGN in galaxy clusters (\lum~$\approx 10^{20}-10^{26}$~W/Hz; \citealt{allen06,rafferty06}) \citet{merloni07} derived a relationship between the kinetic and (5~GHz) radio core luminosities, corrected for Doppler boosting, finding a slope of $A=0.81$ (with a scatter of $\sigma\approx0.5$~dex). The high-luminosity end of this relation was supplemented using a sample of powerful FR~II galaxies ($z=0.056-1.8$; \lum~$>10^{27}$~\wh ) by \citet{daly12}. They find a slope of $A=0.84$ and no evidence for a redshift evolution of this relation.

Recently the relations based on the data used in \citet{birzan04}, \citet{ cavagnolo10}, and \citet{osullivan11} were reassessed by \citet{godfrey16} who have found only a weak correlation with slope $A=0.27$ between the two quantities when distance to the systems in the sample (i.e., Malmquist bias) is taken into account (see \eq{eq:scl} \  and their Tab.~2).\footnote{The data used by the authors are constrained only out to a luminosity distance of $d_L(z)$=100~Mpc (z=0.23).} However, contrary to the inferred flatter slope of the distant-dependent relation, based on theoretical grounds \citet{godfrey16} have predicted a much steeper slope ($A\gtrsim0.5$), which is consistent with those inferred by all of the abovementioned studies. These authors have attributed this discrepancy to either additional model parameters being correlated with the radio jet luminosity or a systematic bias in the X-ray cavity luminosity calculation that induces a flattening in the observed slope. 

Further evidence for a steeper slope ($A\approx0.8-0.9$) of the scaling relation arises from the analytic derivation by \citet{willott99}. \citet{willott99} estimated the kinetic energy using minimum energy arguments, i.e.,\ computing the minimum  energy stored in the lobes to produce the observed synchrotron luminosity, considering the age of the source and the efficiency of conversion of the kinetic luminosity into the internal energy of the observed synchrotron emission. These authors inferred a slope of the relation of $A=0.86$. \citet{willott99} used a reference frequency of 151~MHz, and they folded all uncertainty factors in the calculation (such as departure from minimum energy conditions, uncertainty in the energy in nonradiating particles, and the composition of the jet) into one parameter, $f_\mathrm{W}$, estimated to lie in the range of $f_\mathrm{W}\approx1-20$. We here adopted the conversion to a reference frequency of 1.4~GHz as presented by \citet{heckman14} (their eq.~1), noting that conversions using single power-law spectrum assumptions add additional uncertainties (see also below). 
 
 The various scaling relations, shown in \f{fig:pcav} , can be summarized as 
\begin{equation}
\label{eq:scl}
\log{L_\mathrm{kin}}= A\cdot\log{  \frac{L_\mathrm{1.4GHz }}{10^{24}~\mathrm{W/Hz}}    } + B
,\end{equation}

where $L_\mathrm{kin}$ is the kinetic luminosity in units of W,  $L_\mathrm{1.4GHz}$ is the monochromatic radio luminosity at rest-frame 1.4~GHz in units of W/Hz, and A and B are constants as determined by the various authors as follows:

{\footnotesize
$$
(A,B) = 
\left\{
\begin{array}{l@{,\ }l}
(0.86, 34.72 + 1.5\log{f_\mathrm{W}}) & \mathrm{Willott\, et\, al.\, (1999)}\\
(0.81, 37.39) & \mathrm{ Merloni \,\&\, Heinz\, (2007)}\\
(0.75, 37.02) & \mathrm{ Cavagnolo\, et\, al.\, (2010)}\\
(0.63, 36.76) & \mathrm{ O'Sullivan\, et\, al.\, (2011)}\\
(0.84, 35.69) & \mathrm{ Daly\, et\, al.\, (2012)}\\
(0.27, 36.56) + 1.4\log{\frac{d_L}{100}}) & \mathrm{ Godfrey \,\&\, Shabala\, (2016)},
\end{array} \right.
$$
 }
 where $f_\mathrm{W}$ is the factor defined by \citet{willott99} encompassing all the possible systematic erros ($f_\mathrm{W}\approx1-20$) and $d_\mathrm{L}$ is the luminosity distance expressed in units of Mpc. 

 \citet{merloni07} originally based their scaling relation using 5~GHz core luminosities in units of W (i.e., $L_\nu^\mathrm{core}=\nu\cdot L_\mathrm{nu}^\mathrm{core}$, where $\nu=5$~GHz, and $L_\mathrm{nu}^\mathrm{core}$ is in units of W/Hz). We scaled  this relation in such a way that the total rest-frame 1.4~GHz luminosity of the source is taken as input. This was carried out by converting between rest-frame 5 and 1.4~GHz luminosity using a spectral index of $\alpha=-0.5$ \citep{kimball08} and assuming that the core luminosity on average corresponds to half of the total luminosity of the AGN as supported by our VLA-COSMOS radio AGN also detected by the VLBA-COSMOS survey at $0.005\arcsec$ angular resolution (Herrera Ruiz et al., in prep.). 
We caution, however, that such a conversion may not be applicable at the high-luminosity end (for sources such as, e.g., Cygnus A, not sampled well within the 2 square degreee COSMOS field), and that the conversion, thus, may be luminosity dependent, which is  not accounted for here.
 Similarly, the original expression derived by \citet{daly12} is based on total 178~MHz luminosities converted here to monochromatic 1.4~GHz luminosities using a spectral index of $\alpha=-0.7$.

\begin{figure}
\includegraphics[width=\columnwidth]{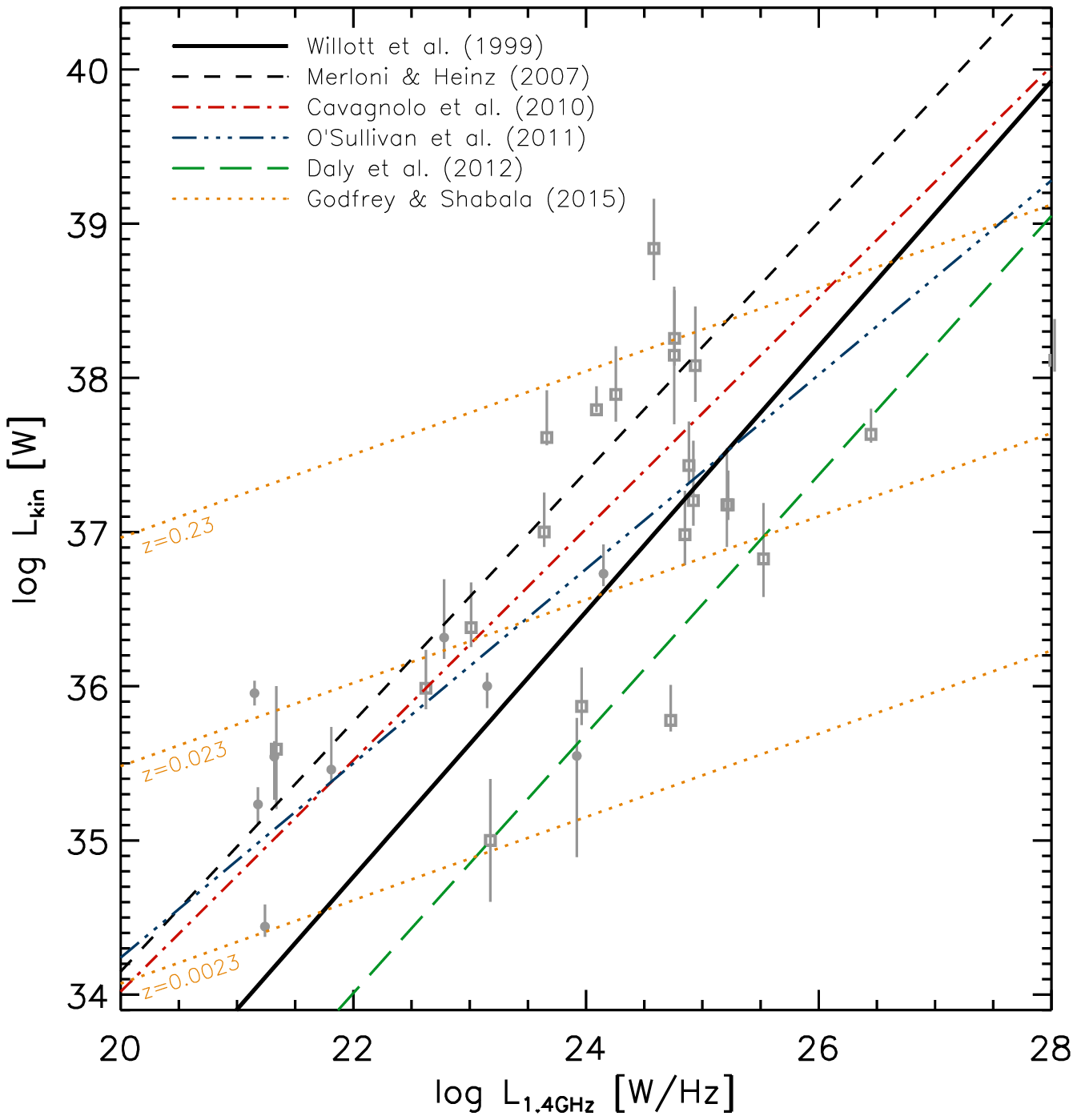}
\caption{
Various scaling relations between monochromatic 1.4~GHz luminosity (\lum ) and kinetic luminosity ($L_\mathrm{kin}$) from the literature, as indicated in the top left of the panel and described in detail in the text (see \eq{eq:scl} ).   Also shown, to guide the eye, are the data used by \citet{birzan04} and \citet{osullivan11} (\lum~$\approx10^{20}-10^{28}$~\wh ; filled symbols). 
The analytically derived \citet{willott99} relation shown is that for $f_\mathrm{W}=15$ (see text for details) in agreement with the relations based on a combination of X-ray cluster/group and radio data (symbols). For $f_\mathrm{W}=4,$ it would  agree with the  \citet{daly12} relation inferred for powerful radio galaxies (\lum$>10^{27}$~\wh ).
  \label{fig:pcav}}
\end{figure}

In \f{fig:lkinapp} \ we show the  kinetic luminosity density as a function of redshift for our pure luminosity or density evolution two-parameter models and using the various scaling relations. Given the similar ($A\approx0.8$) slope of most of the relations the resulting kinetic luminosity density approximately only systematically shifts along the $y$-axis. The \citet{willott99} relation taking $f_W=1$ represents a robust lower limit, and, when considering only the $A=0.6-0.9$ slope relations, the highest values are obtained using the \citet{merloni07} relation. 

The result based on the \citet{godfrey16} relation is not shown in \f{fig:lkinapp}
 \ as it is dependent on the lower integration boundary given the shallow slope of the relation ($A=0.27$). Furthermore, it is distance dependent and constrained only based on a sample  out to $z\leq0.23$. Thus, extrapolating this relation beyond $z\sim0.2$ and up to $z\sim5$ would yield extremely high kinetic luminosity values. For example, 
 for \lum~$=10^{25}$~\wh \ the extrapolated \citet{godfrey16} relation would yield
 $L_\mathrm{kin}=(0.8, 1.6, 2.5, 3.7)\cdot10^{40}$~W for $z=2,3,4,5$, respectively, while the maximum kinetic luminosity for the same input parameters based on the other relations would be $L_\mathrm{kin}=1.6\cdot10^{38}$~W (\citealt{merloni07} relation), thus about two order of magnitude lower. However, out to $z\sim0.1$ the  \citet{godfrey16} parametrization yields results that are consistent with the range of those obtained using the other relations considered, and at $z>0.3$ this parametrization results in kinetic luminosity densities that are increasingly higher by more than an order of magnitude compared to the results based on the other relations.

\begin{figure}
\includegraphics[bb= 0 0 550 339, width=\columnwidth]{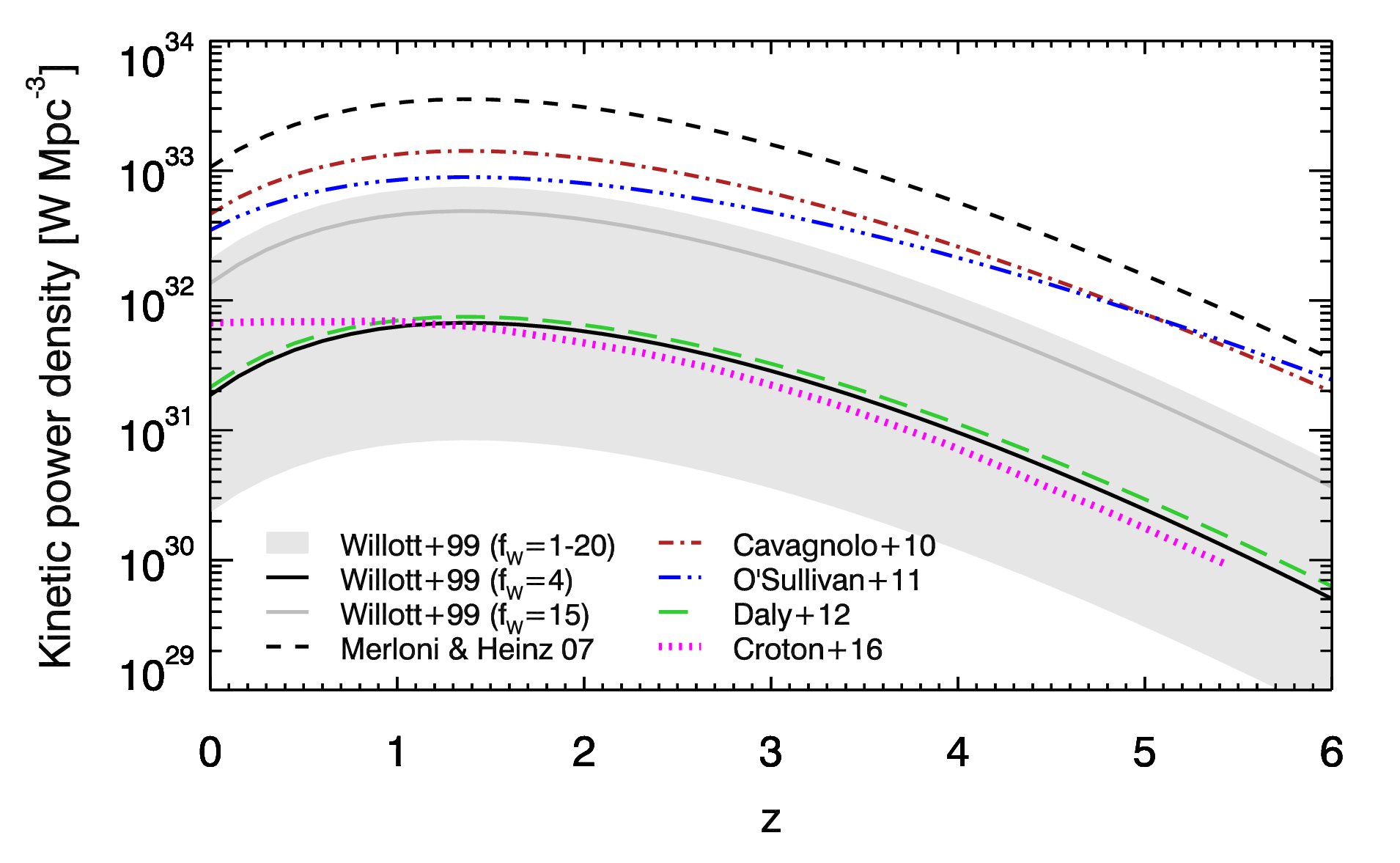}
\includegraphics[bb= 0 0 550 339, width=\columnwidth]{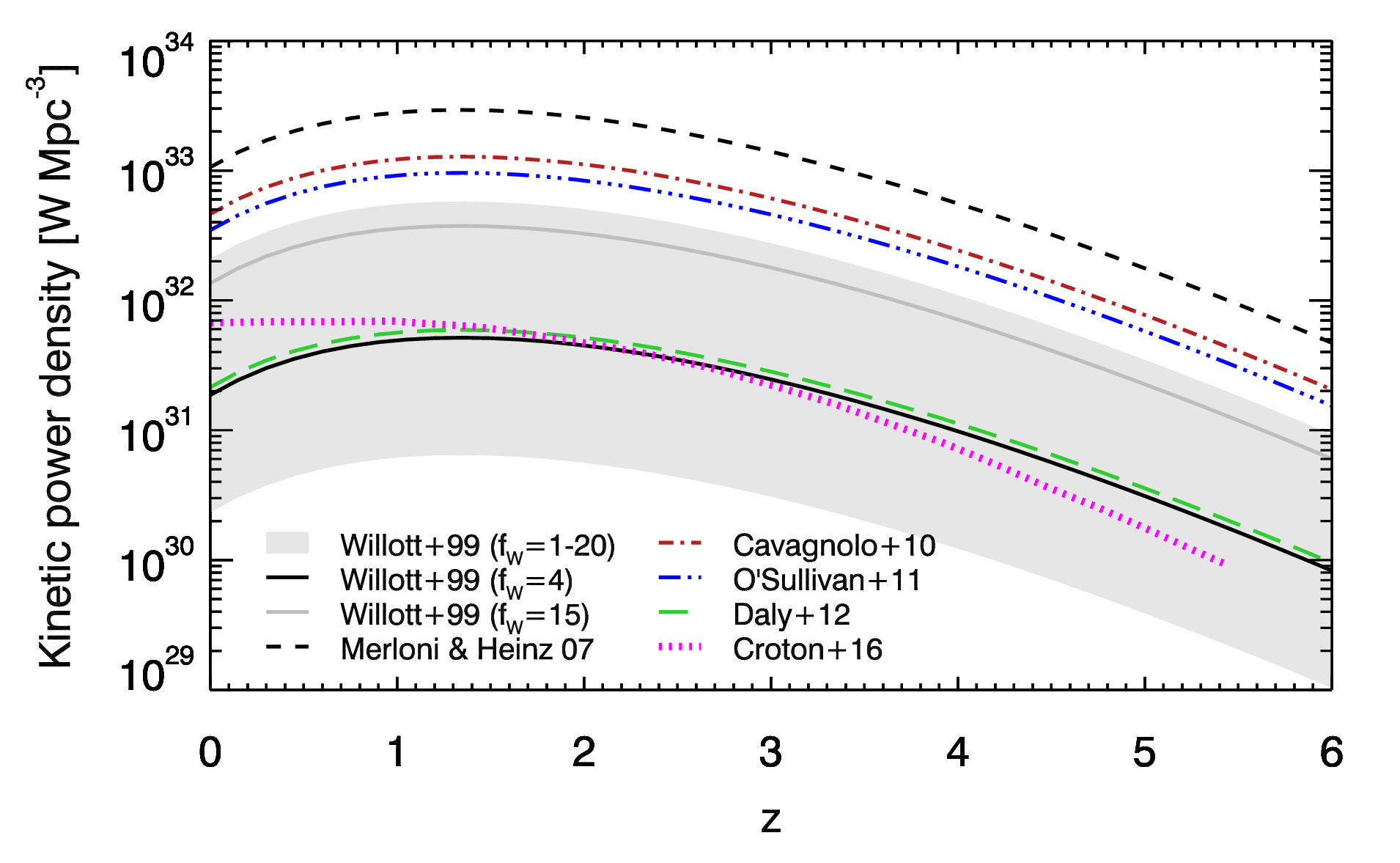}
\caption{Top (bottom) panel shows the kinetic luminosity density as a function of redshift for our pure luminosity (density) evolution 2-parameter model with the various scaling relations commonly applied in the literature, as indicated in each panel.  For comparison the radio model related SMBH accretion luminosity from the SAGE semi-analytic model \citep{croton16} is also shown.
  \label{fig:lkinapp}}
\end{figure}

\end{document}